\documentclass[11pt,fleqn]{article}

\usepackage{a4}
\usepackage{amstext}
\usepackage{amsfonts}
\usepackage{amssymb}
\usepackage{booktabs}
\usepackage{color}
\usepackage{caption}         
\usepackage{cite}
\usepackage{dsfont}
\usepackage{epsfig}
\usepackage{mathtools}
\usepackage{multirow}
\usepackage{physics}
\usepackage{ulem}
\usepackage{subcaption}   
\usepackage{hyperref}
\usepackage{cases}
\usepackage{array}

\usepackage[
linecolor=orange!80!white, backgroundcolor=orange!10!white,
bordercolor=red, textsize=small,]{todonotes}

\newcommand{\gtapprox}{\raisebox{-0.5ex}{$\,\stackrel{>}{\scriptstyle\sim}\,$}}
\newcommand{\ltapprox}{\raisebox{-0.5ex}{$\,\stackrel{<}{\scriptstyle\sim}\,$}}
\newcommand{\RN}[1]{\uppercase\expandafter{\romannumeral#1}}

\begin{document}


	\begin{center}
		
		{\huge \bf Hybrid static potentials in SU(3) lattice gauge theory at small quark-antiquark separations}

		\vspace{0.5cm}
		
		\textbf{Carolin Schlosser$^{a,b}$, Marc Wagner$^{a,b}$}
		
		$^a$~Goethe-Universit\"at Frankfurt am Main, Institut f\"ur Theoretische Physik, Max-von-Laue-Stra{\ss}e 1, D-60438 Frankfurt am Main, Germany \\
		$^b$~Helmholtz Research Academy Hesse for FAIR, Campus Riedberg, Max-von-Laue-Stra{\ss}e 12, D-60438 Frankfurt am Main, Germany
		
		\vspace{0.5cm}
		
		\today
		
	\end{center}
	
	\begin{tabular*}{16cm}{l@{\extracolsep{\fill}}r} \hline \end{tabular*}
	
	\vspace{-0.4cm}
	\begin{center} \textbf{Abstract} \end{center}
	\vspace{-0.4cm}
	We compute the $\Pi_u$ and $\Sigma_u^-$ hybrid static potentials in
	$\text{SU(3)}$ lattice gauge theory using four different lattice spacings ranging from $a = 0.040\,\text{fm}$ to $a = 0.093\,\text{fm}$. 
	We provide lattice data points for quark-antiquark separations as small as $0.08\, \text{fm}$, where the $a$-dependent self-energy as well as lattice discretization errors at tree-level of perturbation theory and at leading order in $a^2$ have been removed.
	We also investigate and exclude possibly present systematic errors from topological freezing, due to the finite spatial lattice volume and from glueball decays.
	Moreover, we provide corresponding parametrizations of the potentials, which can e.g.\ be used for Born-Oppenheimer predictions of heavy hybrid mesons.
	
	\begin{tabular*}{16cm}{l@{\extracolsep{\fill}}r} \hline \end{tabular*}
	
	\thispagestyle{empty}


	\newpage
	
	\setcounter{page}{1}

	
	\section{Introduction}

	The constituent quark model is quite successful in explaining the properties of a variety of non-exotic hadrons, quark-antiquark pairs or triplets of quarks or antiquarks without gluonic excitations.
	However, a particular class of exotic mesons, so-called hybrid mesons, contain such gluonic excitations and, thus, cannot be studied in a proper way using the constituent quark model. These systems require approaches closer to QCD, which contain gluons as degrees of freedom.
	In this work we use lattice gauge theory and are interested in heavy hybrid mesons, which are composed of heavy $c$ or $b$ quarks and a surrounding excited gluon field.
	The gluonic excitation contributes to the quantum numbers of the hybrid meson such that exotic combinations of $J^{PC}$ are allowed, which do not exist in the constituent quark model.
	
	The experimental search for exotic states in existing and future facilities like the GlueX experiment at Jefferson Lab or the PANDA experiment at FAIR as well as the theoretical explanation of their internal structure and properties are currently hot research topics (for an experimental review see e.g.\ Ref.\ \cite{Olsen:2017bmm}, for theoretical reports we refer to Refs.\ \cite{Braaten:2014ita,Meyer:2015eta,Swanson:2015wgq,Lebed:2016hpi,Brambilla:2019esw}).
	Concerning theoretical approaches, lattice gauge theory is an ideal non-perturbative first principles approach to investigate properties and masses of heavy hybrid mesons, either within the Born-Oppenheimer approximation~ \cite{Perantonis:1990dy,Juge:1997nc,Juge:1999ie,Guo:2008yz,Braaten:2014qka,Berwein:2015vca,Oncala:2017hop,Capitani:2018rox,Brambilla:2018pyn,Brambilla:2019jfi} or in full lattice QCD (see e.g.\  Refs.\ \cite{Bernard:2003jd,Bali:2011rd,Cheung:2016bym,Ray:2021nhe}).
	We focus on hybrid mesons composed of heavy $c$ or $b$ quarks and use $\text{SU(3)}$ lattice gauge theory in combination with the Born-Oppenheimer approximation \cite{Born:1927}, which is a two-step approach.
	In the first step, we fix the positions of the heavy quarks and compute so-called hybrid static potentials with lattice gauge theory. Hybrid static potentials correspond to energy levels of gluonic excitations in the presence of static quarks as functions of their separation.
	In the second step of the Born-Oppenheimer approximation, the radial Schrödinger equation for the relative coordinate of the heavy quark-antiquark pair is solved with one of the hybrid static potentials obtained in the first step.
	
	In recent years a lot of effort was invested to refine the second step of the Born-Oppenheimer approximation, e.g.\ by including the mixing of different sectors via coupled channel equations~\cite{Braaten:2014qka,Berwein:2015vca,Oncala:2017hop} and by taking heavy quark spin effects into account~\cite{Brambilla:2018pyn,Brambilla:2019jfi}.
	These approaches require precise lattice results for hybrid static potentials, in particular at small quark-antiquark separations $r$ to combine them with perturbative predictions valid only at small $r$ or to fix matching coefficients in potential Non-Relativistic QCD (pNRQCD) \cite{Berwein:2015vca,Brambilla:2017uyf,Brambilla:2018pyn,Brambilla:2019jfi}. Thus, the main goal of this work is to use lattice gauge theory to investigate the small-$r$ region of the $\Pi_u$ and $\Sigma_u^-$ hybrid static potentials. We aim at extending the range of precise lattice field theory results to smaller quark-antiquark separations and improve existing investigations of hybrid static potentials \cite{Griffiths:1983ah,Campbell:1984fe,Campbell:1987nv,Perantonis1989StaticPF,
		Michael:1990az,Perantonis:1990dy,Juge:1997nc,Peardon:1997jr,
		Juge:1997ir,Morningstar:1998xh,Michael:1998tr,Michael:1999ge,
		Juge:1999ie,Juge:1999aw,Bali:2000vr,
		Morningstar:2001nu,Juge:2002br,Juge:2003qd,Michael:2003ai,
		Michael:2003xg,Bali:2003jq,Juge:2003ge,
		Wolf:2014tta,Reisinger:2017btr,Bicudo:2018yhk,Bicudo:2018jbb,Reisinger:2018lne,Capitani:2018rox}.
	For this we perform computations at four different lattice spacings ranging from $a=0.040 \,\text{fm}$ to $a=0.093 \,\text{fm}$.
	These computations at several small lattice spacings do not only allow to access smaller quark-antiquark separations than before, but also to explore and remove lattice discretization errors, such that our final results are expected to be consistent with the continuum limit within statistical errors. Moreover, we can convincingly confirm the repulsive behavior of hybrid static potentials at small $r$ predicted by perturbation theory.
	
	To compute the ordinary static potential and the $\Pi_u$ and $\Sigma_u^-$ hybrid static potentials we employ optimized operators from our previous work \cite{Capitani:2018rox} as well as a multilevel algorithm \cite{Luscher:2001up}. In this way we obtain precise lattice results for these potentials on four ensembles for quark-antiquark separations as small as $0.08 \, \text{fm}$ (see Section~\ref{sec:theory_hybridstaticpotentials} to Section~\ref{sec:latticeresults}).
	We also check that our lattice gauge theory computations are not contaminated by sizable systematic errors related to topological freezing, the finite spatial lattice volume or glueball decays of hybrid flux tubes, which are expected to be particularly prominent at small lattice spacings and small quark-antiquark separations (see Section~\ref{sec:systematicerrors}).
	Moreover, we provide parametrizations describing the hybrid static potentials for quark-antiquark separations $0.08\,\text{fm} \ltapprox r \ltapprox 1.12\,\text{fm}$. We use these parameterizations to eliminate discretization errors and the $a$-dependent self-energy
	(see Section~\ref{sec:parametrization}).
	We also use the parametrizations to check the impact of including our new lattice data at small lattice spacings in Born-Oppenheimer predictions of $c \bar{c}$ and $b \bar{b}$ hybrid meson masses and find sizable differences to our previous work \cite{Capitani:2018rox}, where we have only considered a single lattice spacing $a = 0.093 \, \text{fm}$.
	The numerical values of all lattice data points and their parametrizations are provided for straightforward use in future applications, e.g.\ for predictions of heavy hybrid meson masses in more refined Born-Oppenheimer approaches as proposed in Refs.\ \cite{Berwein:2015vca,Oncala:2017hop,Brambilla:2018pyn,Brambilla:2019jfi}. We also provide similar results for gauge group $\text{SU(2)}$, which were obtained at an early stage of this work.

	\section{\label{sec:theory_hybridstaticpotentials}Hybrid static potentials: quantum numbers, operators and correlation functions}

	Hybrid static potentials represent the energy of the excited gluon field in the presence of a static quark and antiquark as a function of their separation.
	
	Static potentials are characterized by the following three quantum numbers:
	\begin{itemize}
		\item $\Lambda = \Sigma (=0), \Pi (=1), \Delta (=2), \ldots$ denotes the total angular momentum with respect to the quark-antiquark separation axis, i.e.\ is a non-negative integer (w.l.o.g.\ we separate the static quark and antiquark along the $z$-axis).
		\item $\eta = g (=+), u (=-)$ describes the even ($g$) or odd ($u$) behavior under the combined parity and charge conjugation transformation $\mathcal{P} \circ \mathcal{C}$.
		\item $\epsilon = +,-$ is the eigenvalue of a reflection $\mathcal{P}_x$ along an axis perpendicular to the quark-antiquark separation axis (for definiteness we use the $x$-axis).
		For $\Lambda \ge 1$, hybrid static potentials are degenerate with respect to $\epsilon$ and $\epsilon$ is typically omitted.
	\end{itemize}
	The ordinary static potential has quantum numbers $\Sigma_g^+$, while hybrid static potentials have quantum numbers different from $\Sigma_g^+$.
	In this work we carry out a precise computation and parametrization of the two lowest hybrid static potentials, which have quantum numbers $\Pi_u$ and $\Sigma_u^-$, with particular focus on rather small quark-antiquark separations $r$.
	
	Hybrid static potentials are computed from correlation functions similar to Wilson loops, where the straight spatial parallel transporters are replaced by more complicated gauge link combinations with non-trivial transformation properties,
	\begin{eqnarray}\label{eq:CorrelationfunctionW}
		\nonumber & & \hspace{-0.7cm} W_{S;\Lambda_{\eta}^{\epsilon}}(r,t) = \\
		& & = \expval{
			\Tr\left(
			a_{S;\Lambda_{\eta}^{\epsilon}}(-r/2,+r/2;0) U(+r/2;0,t) \left(a_{S;\Lambda_{\eta}^{\epsilon}}(-r/2,+r/2;t)\right)^{\dagger} U(-r/2;t,0)
			\right)
		}_U .
	\end{eqnarray}
	$U(r;t_1,t_2)$ is a straight path of temporal gauge links from time $t_1$ to time $t_2$ at spatial position $\mathbf{r} = (0,0,r)$ and $\expval{\ldots}_U$ denotes the average on an ensemble of gauge link configurations.
	$a_{S;\Lambda_{\eta}^{\epsilon}}$ is given by a sum of properly transformed spatial insertions $USU$, to probe the sector with quantum numbers $\Lambda_{\eta}^{\epsilon}$,
	\begin{eqnarray}
		\nonumber & & \hspace{-0.7cm} a_{S; \Lambda_{\eta}^{\epsilon}} (-r/2, +r/2) = \frac{1}{4} \sum_{k=0}^{3} \exp(\frac{i\pi \Lambda k}{2}) R\left( \frac{\pi k}{2}\right) \\
		\nonumber & & \hspace{0.675cm} \Big(U(-r/2,r_1) \Big(S(r_1,r_2) + \epsilon S_{\mathcal{P}_x}(r_1,r_2)\Big) U(r_2,+r/2) \\
		\label{eq:a} & & \hspace{0.675cm} +	U(-r/2,-r_2) \Big(\eta S_{\mathcal{P} \circ \mathcal{C}}(-r_2,-r_1) + \eta \epsilon S_{(\mathcal{P} \circ \mathcal{C}) \mathcal{P}_x}(-r_2,-r_1)\Big) U(-r_1,+r/2)\Big).
	\end{eqnarray}
	The notation is explained in detail in Ref.\ \cite{Capitani:2018rox}. We employ operators $S$ from Ref.\ \cite{Capitani:2018rox}, where we have carried out a dedicated optimization to maximize the generated ground state overlaps. For the $\Pi_u$ hybrid static potential we use $S_{\RN{3},1}$ and for the $\Sigma_u^-$ hybrid static potential we use $S_{\RN{4},2}$. Detailed definitions can be found in Table~$3$ and Table~$5$ of Ref.\ \cite{Capitani:2018rox}. The operator extents in these tables are given in units of the lattice spacing for $a = 0.093 \, \text{fm}$. For computations at smaller values of $a$ we increase the operator extents in units of the lattice spacing such that they are approximately constant in physical units.
	
	To further enhance the ground state overlaps, we apply APE smearing  to the spatial gauge links appearing in $a_{S;\Lambda_{\eta}^{\epsilon}}$. The number of APE smearing steps is increased with decreasing lattice spacing to keep the smearing radius approximately constant in physical units.
	Details can be found in Appendix~\ref{Appendix:optimization}.

	\section{Computational details}

	\subsection{Gauge link ensembles}\label{sec:latticesetups}
	
		We computed hybrid static potentials both on $\text{SU(2)}$ and $\text{SU(3)}$ gauge link configurations generated with the standard Wilson plaquette action without dynamical quarks.
		Results for purely gluonic observables such as energies in the presence of a static quark-antiquark pair, possibly in a sector with hybrid quantum numbers, are expected to be similar in pure gauge theory and in QCD (for hybrid static potentials this is supported by lattice results from Ref.\ \cite{Bali:2000vr}). To study hybrid static potentials, it might even be advantageous to use pure gauge theory, because in that case an excited flux tube can only decay into multiparticle states, which include rather heavy glueballs, but not light pions. In Section~\ref{sec:glueballdecay} glueball decays are discussed in detail.

		In the main part of this work we focus exclusively on computations and results for gauge group $\text{SU(3)}$. Corresponding results for gauge group $\text{SU(2)}$ are summarized in Appendix~\ref{Appendix:SU2results}.

		We generated four ensembles of gauge link configurations with gauge couplings \\ $\beta = 6.594 \, , \, 6.451 \, , \, 6.284 \, , \, 6.000$ using the CL2QCD software package\cite{Philipsen:2014mra}.
		We relate the corresponding lattice spacing $a$ to the Sommer scale $r_0$ via a parametrization of $\ln(a/r_0)$ provided in Ref.\ \cite{Necco:2001xg}, which is based on a precision determination of $r_0$ up to $\beta=6.92$.
		We introduce physical units by setting $r_0 = 0.5 \, \text{fm}$, which is a simple and common choice in pure gauge theory, but is slightly larger than QCD results \cite{Sommer:2014mea}.
		
		The details of our gauge link ensembles, which we label by $A$, $B$, $C$ and $D$, are collected in Table~\ref{tab:latticesetups4}.
		The lattice volume for all four ensembles is $L^3 \times T \approx (1.2 \, \text{fm})^3 \times 2.4 \,\text{fm}$. This is sufficiently large to neglect finite volume corrections (see Section~\ref{sec:finitevolume} for a detailed investigation and discussion).
		Each ensemble was generated by $N_{\text{sim}}$ independent Monte Carlo simulations, where each simulation comprises $N_{\text{total}}$ updates. An update is composed of a heatbath sweep and $N_{\text{or}}$ overrelaxation sweeps. $N_{\text{or}}$ is chosen roughly as $N_{\text{or}} \approx 1.5 \, r_0 / a$ following Ref.\ \cite{Guagnelli:1998ud}.
		This is expected to minimize correlations between subsequent gauge link configurations.
		The first $N_{\text{therm}}$ updates are considered as thermalization updates and the corresponding gauge link configurations were, thus, discarded. After thermalization, gauge link configurations separated by $N_{\text{sep}}$ updates were used to measure correlation functions. The total number of gauge link configurations used for measurements is thus $N_{\text{meas}} = N_{\text{sim}} (N_{\text{total}} - N_{\text{therm}}) / N_{\text{sep}}$.

		To eliminate autocorrelations, we combined these $N_{\text{meas}}$ gauge link configurations into a much smaller number of bins. Statistical errors were determined using both the jackknife and the bootstrap method. Further details concerning data analysis are discussed in Appendix~\ref{APP012}.

		\begin{table}[htb]
			\begin{center}
				\def\arraystretch{1.2}
				\begin{tabular}{cccccccccc}
					\hline
					ensemble & $\beta$ & $a$ in $\text{fm}$ \cite{Necco:2001xg} & $(L/a)^3 \times T/a$ & $N_{\text{sim}}$ & $N_{\text{total}}$ & $N_{\text{or}}$ & $N_{\text{therm}}$ & $N_{\text{sep}}$ & $N_{\text{meas}}$ \\
					\hline
					$A$ & $6.000$ & $0.093$ & $12^3\times 26$ & $2$ & 
					$60 000$ & $\phantom{0}4$ & $20 000$ & $\phantom{0}50$ & $1600$ \\
					$B$ & $6.284$ & $0.060$ & $20^3\times 40$ & $2$ & 
					$60 000$ & $12$ & $20 000$ & $100$ & $\phantom{0}800$ \\
					$C$ & $6.451$ & $0.048$ & $26^3\times 50$ & $4$ & 
					$80 000$ & $15$ & $40 000$ & $200$ & $\phantom{0}800$ \\ 
					$D$ & $6.594$ & $0.040$ & $30^3\times 60$ & $4$ & 
					$80 000$ & $15$ & $40 000$ & $200$ & $\phantom{0}800$ \\
					\hline
				\end{tabular}
			\end{center}
			\caption{Gauge link ensembles.}
			\label{tab:latticesetups4}
		\end{table}

	\subsection{Multilevel algorithm}\label{sec:multilevelalgorithm}
	
		For the efficient computation of Wilson loop-like correlation functions (\ref{eq:CorrelationfunctionW}) we employ the multilevel algorithm \cite{Luscher:2001up}.
		The starting point for a multilevel simulation is one of the sets of $N_{\text{meas}}$ thermalized gauge link configurations discussed in Section~\ref{sec:latticesetups}.
		The lattice is partitioned into $n_{\text{ts}}$ time-slices with thicknesses $p_1,p_2, \dots, p_{n_{\text{ts}}}$.
		In principle, time-slices can be partitioned more than once, but we use only a single level of partitioning.
		For each time-slice $n_m$ sublattice configurations are generated using a standard heatbath algorithm. These sublattice configurations are separated by $n_u$ heatbath sweeps, where links in the interior of the time-slice are updated, while spatial links on the boundaries are fixed.

		Two-link operators are defined via $\mathds{T}(x,r \hat{j})_{\alpha \beta \gamma \delta} = U_0^*(x)_{\alpha\beta} U_0(x+r \hat{j})_{\gamma\delta}$ ($\hat{j}$ denotes the spatial unit vector in $j$-direction, e.g.\ $\hat{1} = (0,1,0,0)$). They are multiplied according to
		\begin{equation}
			 \mathds{P}_k = \{\mathds{T}(x+(d_k-p_k) a \hat{0},r \hat{j}) \mathds{T}(x+(d_k-p_k+1) a \hat{0},r \hat{j}) \dots \mathds{T}(x+(d_k-1) a \hat{0},r \hat{j})\}
		\end{equation}
		with the multiplication prescription $\{\mathds{T}_1 \mathds{T}_2\}_{\alpha \beta \gamma \delta} = \{\mathds{T}_1\}_{\alpha \sigma \gamma \rho} \{\mathds{T}_2\}_{\sigma \beta \rho \delta}$, i.e.\ such that the product $\mathds{P}_k$ connects the two boundaries of the time-slice $k$, i.e.\ extends from $t/a = d_{k-1}$ to $t/a = d_k$ with $d_k = \sum_{j=1}^k p_j$. The products $\mathds{P}_k$ are then averaged over the $n_m$ corresponding sublattice configurations with the results denoted as $[\mathds{P}_k]$.
		
		Wilson loops are computed via
		\begin{equation}
			\label{EQN854} W_{S;\Lambda_\eta^\epsilon}(r,t) = a_{S;\Lambda_\eta^\epsilon}(\mathbf{x},\mathbf{x}+r \hat{j};x_0)_{\alpha \gamma} \{[\mathds{P}_{k}] [\mathds{P}_{k+1}] \dots [\mathds{P}_{k+n_t-1}]\}_{\alpha \beta \gamma \delta} \Big(a_{S;\Lambda_\eta^\epsilon}(\mathbf{x},\mathbf{x}+r \hat{j};x_0+t)\Big)^\ast_{\beta \delta} ,
		\end{equation}
		where the spatial parallel transporters $a_{S;\Lambda_\eta^\epsilon}$ (see Eq.\ (\ref{eq:a})) are both located on boundaries between time-slices.
		$n_t$ denotes the number of time-slices traversed by the Wilson loop, i.e.\ $d_{k-1} = x_0$ and $\sum_{j=1}^{n_t} p_{k+j} = t/a$. Finally, the samples $W_{S;\Lambda_\eta^\epsilon}(r,t)$ from Eq.\ (\ref{EQN854}) are averaged over space-time, the three spatial directions and the $N_{\text{meas}}$ thermalized gauge link configurations.
		
		Note that the time-slice partitioning might impose constraints on the temporal extent of Wilson loops, which can be computed.
		For simplicity we choose a regular pattern, where all time-slices have thickness $2$, i.e.\ $p_1 = p_2 = \dots = p_{n_{\text{ts}}} = 2$.
		This choice is not only simple but also efficient, because it allows to exploit translational invariance in temporal direction extensively.
		Moreover, we use $n_m = 400$ and $n_u = 30$.

		For a technically more detailed discussion of the multilevel algorithm see Section~3.2 of Ref.\ \cite{Brambilla:2021wqs}.

	\subsection{Tree-level improvement for static potentials}\label{sec:treelevelimprovement}
	
		To reduce lattice discretization errors for the ordinary and for hybrid static potentials, we apply a tree-level improvement similar as in Ref.\ \cite{Hasenfratz:2001tw}.
		In the continuum in leading-order perturbation theory static potentials are proportional to $1/r$ due to one-gluon-exchange. 
		The ordinary static potential is attractive, while the $\Pi_u$ and $\Sigma_u^-$ hybrid static potentials exhibit a repulsive $1/r$ behavior, which is suppressed by the factor $1/8$.
		On an infinite spacetime lattice the leading order perturbative result can be computed in a straightforward way as discussed in Appendix~\ref{Appendix:treelevelimprovement}. The difference to its $1/r$ continuum counterpart represents lattice discretization errors at tree-level. 
		These discretization errors can be subtracted from the non-perturbative lattice data points obtained from Wilson loop-like correlation functions (\ref{eq:CorrelationfunctionW}). 
		For this one needs to estimate the prefactor of the $1/r$ perturbative part, which is proportional to the strong coupling (see e.g.\ Ref.\ \cite{Jansen:2011vv} and references therein). 
		We do this in Section~\ref{sec:parametrization} with a suitable fit to the $\Sigma_g^+$ static potential.
	
		We note that there is a related but slightly different method for tree-level improvement also common in the literature (see e.g.\ Refs.\ \cite{Necco:2001xg,Jansen:2011vv}). Instead of changing the lattice result for the value of the static potential at a given quark-antiquark separation, this separation is replaced by a so-called improved separation. 
		According to our numerical tests this method works well for the static force. 
		However, for static potentials it seems to be inferior to the method discussed in the previous paragraph, because of their linear behavior for large separations. 
		We plan to discuss this in detail in another publication.

	\section{Lattice field theory results for the $\Pi_u$ and $\Sigma_u^-$ hybrid static potentials}\label{sec:latticeresults}

	In the following we discuss our lattice field theory results $V^e_{\Lambda_\eta^\epsilon}(r)$ for static potentials with quantum numbers $\Lambda_\eta^\epsilon = \Sigma_g^+$ (the ordinary static potential) and $\Lambda_\eta^\epsilon = \Pi_u , \Sigma_u^-$ (the two lowest hybrid static potentials) for all four lattice ensembles $e \in \{A, B, C, D\}$ listed in Table~\ref{tab:latticesetups4}. 
	They correspond to the ground state energies in the sectors with quantum numbers $\Lambda_\eta^\epsilon$ and quark-antiquark separation $r$.
	
	To extract these static potentials, we compute temporal correlation functions $W^e_{\Lambda_\eta^\epsilon}(r,t)$ (see Eq.~\eqref{eq:CorrelationfunctionW}) of suitably designed creation operators as discussed in Section~\ref{sec:theory_hybridstaticpotentials}. 
	We restrict the computations to temporal separations $t$, which are multiples of $2a$. 
	This allows the use of a single multilevel time-slice partitioning, which is simple as well as efficient (for details see Section~\ref{sec:multilevelalgorithm}).
	
	Effective potentials are defined in terms of the correlation functions $W^e_{\Lambda_\eta^\epsilon}(r,t)$ via
	\begin{equation}
		V^e_{\text{eff};\Lambda_{\eta}^{\epsilon}}(r,t) = \frac{1}{2 a} \ln(\frac{W^e_{\Lambda_{\eta}^{\epsilon}}(r,t)}{W^e_{\Lambda_{\eta}^{\epsilon}}(r,t+2a)}) .
	\end{equation}
	These effective potentials approach plateaus at large $t$, which correspond to the ground state energies,	i.e.\ 
	\begin{equation}
		\label{EQN833}
		V^e_{\Lambda_\eta^\epsilon}(r) = \lim_{t \rightarrow \infty} V^e_{\text{eff};\Lambda_{\eta}^{\epsilon}}(r,t) .
	\end{equation}
	
	Numerically the plateau values and, thus, the static potentials are extracted by uncorrelated $\chi^2$ minimizing fits of constants to $a V^e_{\text{eff};\Lambda_\eta^\epsilon}(r,t)$ in the range $t^\prime_\text{min} \le t \le t^\prime_\text{max}$. 
	The fit range is chosen individually for each set of quantum numbers $\Lambda_\eta^\epsilon$ and each quark-antiquark separation $r$ by an algorithm used already in our preceding work \cite{Capitani:2018rox}:
	\begin{itemize}
		\item $t_{\text{min}}$ is defined as the minimal $t$, where the values of $a V^e_{\text{eff};\Lambda_\eta^\epsilon}(r,t-2a)$ and $a V^e_{\text{eff};\Lambda_\eta^\epsilon}(r,t)$ differ by less than $1 \, \sigma$.
		\item $t_{\text{max}}$ is the maximal $t$, where $W^e_{\Lambda_\eta^\epsilon}(r,t+2a)$ has been computed, and where its statistical error is reasonably small.
		\item Fits to $a V^e_{\text{eff};\Lambda_\eta^\epsilon}(r,t)$ are performed for all ranges $t^\prime_\text{min} \le t \le t^\prime_\text{max}$ with $t_\text{min} \le t^\prime_\text{min}$, $t^\prime_\text{max} \le t_\text{max}$ and $t^\prime_\text{max} - t^\prime_\text{min} \ge 6 \, a$.
		\item The result of the fit with the longest plateau and $\chi_\text{red}^2 \le 1$ is taken as result for $V^e_{\Lambda_\eta^\epsilon}(r)$.
	\end{itemize}

\begin{figure}[htb]
		\begin{minipage}{0.5\linewidth}
			\includegraphics[width=\linewidth,page=1]{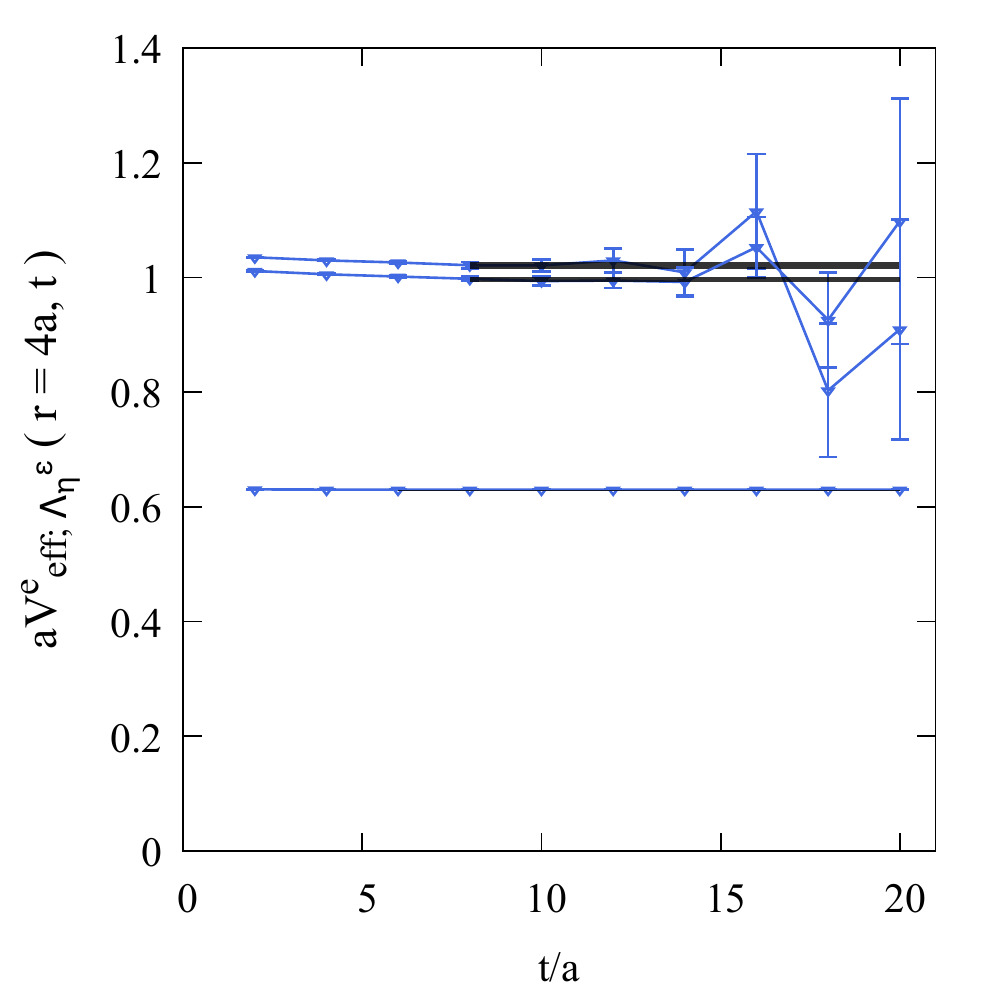}
		\end{minipage}
		\begin{minipage}{0.5\linewidth}
			\includegraphics[width=\linewidth,page=2]{plots/Plot_Veff}
		\end{minipage}
		\caption{Exemplary plots of effective potentials $a V^e_{\text{eff};\Lambda_{\eta}^{\epsilon}}(r,t)$ with $\Lambda_\eta^\epsilon = \Sigma_g^+,\Pi_u , \Sigma_u^-$ for $r= 4a$ (Left: ensemble $B$, i.e.\ $a=0.060 \, \text{fm}$; Right: ensemble $D$, i.e.\ $a=0.040 \, \text{fm}$).}
		\label{fig:effpotentials0}
	\end{figure}

	To illustrate the quality of our lattice data, we show exemplary effective potential plots in Figure~\ref{fig:effpotentials0}. 
	The final fit ranges $t^\prime_\text{min} \le t \le t^\prime_\text{max}$ and fit results are indicated by the horizontal lines.

	The resulting static potentials 
	$V^e_{\Lambda_\eta^\epsilon}(r)$ for $\Lambda_\eta^\epsilon = \Sigma_g^+,\Pi_u , \Sigma_u^-$ and
	separations $r \geq 2 a$ are collected in units of the lattice spacing in Appendix~\ref{Appendix:summarylatticedata}, Table~\ref{tab:latticedata_ABCD_all}.
	Due to the regulator-dependent self-energy of static quarks, potentials computed at different lattice spacings $a$, i.e.\ on different ensembles $e$, are shifted relative to each other. 
	We will subtract these self-energies in Section~\ref{sec:parametrization}, where we also remove discretization errors at tree-level and partly proportional to $a^2$, before we show the results for all ensembles together in a common plot in Figure~\ref{fig:parametrizedlatticedata1} and list them in physical units in Table~\ref{tab:latticedata_ABCD_tilde_all}.
	
	Note that in contrast to previous lattice field theory computations of hybrid static potentials \cite{Perantonis1989StaticPF,Michael:1990az,Perantonis:1990dy,Juge:1997nc,Juge:1997ir,Capitani:2018rox,Juge:2002br,Bali:2003jq,Juge:2003ge}, where lattice spacings $a \gtapprox 0.07 \, \text{fm}$ were used \footnote{In Ref.\ \cite{Michael:1990az} hybrid static potentials were computed for gauge group SU(2) at very small lattice spacing $a \approx 0.022 \, \text{fm}$ (when setting the scale as in Ref.\ \cite{Hirakida:2018uoy}), but at the same time also at very small spatial volume, such that finite volume effects appear to be huge (see e.g.\ Figure~1 in Ref.\ \cite{Michael:1990az} and our our detailed discussion of finite volume effects in Section~\ref{sec:finitevolume}).}, our results are based on four ensembles with lattice spacings as small as $0.04 \, \text{fm}$. 
	Since lattice discretization errors in static potentials typically become large for $r \ltapprox 2 \, a$, the lattice potentials presented in this work are trustworthy down to $r \approx 0.08 \, \text{fm}$, whereas existing works were limited to separations roughly twice as large. 
	
	A major goal of this work is to explore the small-$r$ region of the $\Pi_u$ and $\Sigma_u^-$ hybrid static potentials to make contact to perturbative calculations. 
	Using the framework of potential Non Relativistic QCD (pNRQCD) these hybrid static potentials have been predicted to be repulsive at very small $r$ \cite{Brambilla:1999xf,Bali:2003jq,Berwein:2015vca,Oncala:2017hop,Soto:2017one}, a behavior, which could not convincingly be confirmed by existing lattice computations, because of the use of rather coarse lattice spacings. 
	In contrast to that, our results from the ensembles with the fine lattice spacings $a \approx 0.048 \, \text{fm}$ and $a \approx 0.040 \, \text{fm}$ clearly show the predicted and expected upward curvature at small $r$ (see Table~\ref{tab:latticedata_ABCD_all}, Table~\ref{tab:latticedata_ABCD_tilde_all} and Figure~\ref{fig:parametrizedlatticedata1}). 
	This will be discussed in detail in Section~\ref{sec:parametrization}, where we parametrize our lattice data points by analytic functions based on pNRQCD predictions.

	\section{Parametrization of lattice results for hybrid static potentials} \label{sec:parametrization}

	In this section we parametrize the lattice data points for the ordinary static potential $V_{\Sigma_g^+}(r)$ and the two lowest hybrid static potentials $V_{\Pi_u}(r)$ and $V_{\Sigma_u^-}(r)$ computed in Section~\ref{sec:latticeresults} and collected in Table~\ref{tab:latticedata_ABCD_all}.
	The resulting parametrizations allow to eliminate discretization errors to a large extent and can e.g.\ be used as input for Born-Oppenheimer predictions of heavy hybrid meson masses as previously done in Refs.\ \cite{Capitani:2018rox,Berwein:2015vca,Perantonis:1990dy,Juge:1997nc,Juge:1999ie,Braaten:2014qka,Oncala:2017hop,Guo:2008yz,Brambilla:2018pyn,Brambilla:2019jfi}. 
	
	In addition to the lattice data points specifically computed in the context of this work and discussed in Section~\ref{sec:latticeresults}, we use results from our previous computation \cite{Capitani:2018rox} for quark-antiquark separations $0.19 \, \text{fm} \ltapprox r \ltapprox 1.12 \, \text{fm}$ to constrain our parametrizations also at large separations.
	This computation was performed at lattice spacing $a \approx 0.093\,\text{fm}$, which is identical to the largest lattice spacing used in this work,
	 i.e.\ to that of ensemble $A$.
	However, in contrast to the computations discussed in Section~\ref{sec:latticeresults}, HYP2 smeared temporal links were used, which imply a significantly reduced self-energy and consequently smaller statistical errors, but possibly also larger discretization errors at small separations $r$.
	For completeness, these previous lattice results $V^{A^\text{HYP2}}_{\Lambda_\eta^\epsilon}(r)$ are also listed in Table~\ref{tab:latticedata_ABCD_all}.
	
	When combining the lattice results for the static potentials from the five ensembles \\ $e \in \{ A, B, C, D, A^\text{HYP2} \}$ ($A^\text{HYP2}$ denotes the ensemble generated in the context of Ref.\ \cite{Capitani:2018rox}), one needs to take into account that the self-energy is different for each ensemble. 
	It depends both on the lattice spacing $a$ and whether HYP2 smeared temporal links are used or not.
	To eliminate both the self-energy and lattice discretization errors at tree-level, we first perform an 8-parameter uncorrelated $\chi^2$-minimizing fit of
	\begin{equation}\label{eq:Cornellpotential_fitfunction}
		V_{\Sigma_g^+}^{\text{fit},e}(r) = V_{\Sigma_g^+}(r) + C^e + \Delta V^{\text{lat},e}_{\Sigma_g^+}(r)
	\end{equation}
	with the Cornell ansatz
	\begin{equation}\label{eq:Cornellpotential}
		V_{\Sigma_g^+}(r) = -\frac{\alpha}{r} + \sigma r
	\end{equation}
	and
	\begin{equation}\label{eq:latticeartifacts}
		\Delta V^{\text{lat},e}_{\Sigma_g^+}(r) = \alpha' \bigg(\frac{1}{r} - \frac{G^e(r/a)}{a}\bigg)
	\end{equation}
	to all lattice data points $V_{\Sigma_g^+}^e(r)$ with $0.2 \, \text{fm} < r$. The fit parameters are the $1/r$ coefficient $\alpha$, the string tension $\sigma$, the coefficient $\alpha'$ and for each ensemble an additive constant $C^e$.
	The constants $C^e$ absorb the ensemble dependent self energies. $\Delta V^{\text{lat},e}_{\Sigma_g^+}(r)$ reflects lattice discretization errors at tree level, where the continuum result for the ordinary static potential at tree level is proportional to $1/r$ and its lattice counterpart $G^e(r/a)/a$ can be calculated numerically (see Refs.\ \cite{Luscher:1995zz,Necco:2001xg,Hasenfratz:2001tw}	and Appendix~\ref{Appendix:treelevelimprovement}). 
	The physically meaningful part of the parametrization (\ref{eq:Cornellpotential_fitfunction}) is $V_{\Sigma_g^+}(r)$ with the two parameters $\alpha$ and $\sigma$. 
	It is known that this Cornell ansatz provides an accurate description of the ordinary static potential for $0.2 \, \text{fm} \ltapprox r$ (see e.g.\ Ref.\ \cite{Karbstein:2018mzo}).
	
\begin{figure}[p]\centering
		\includegraphics[width=\linewidth]{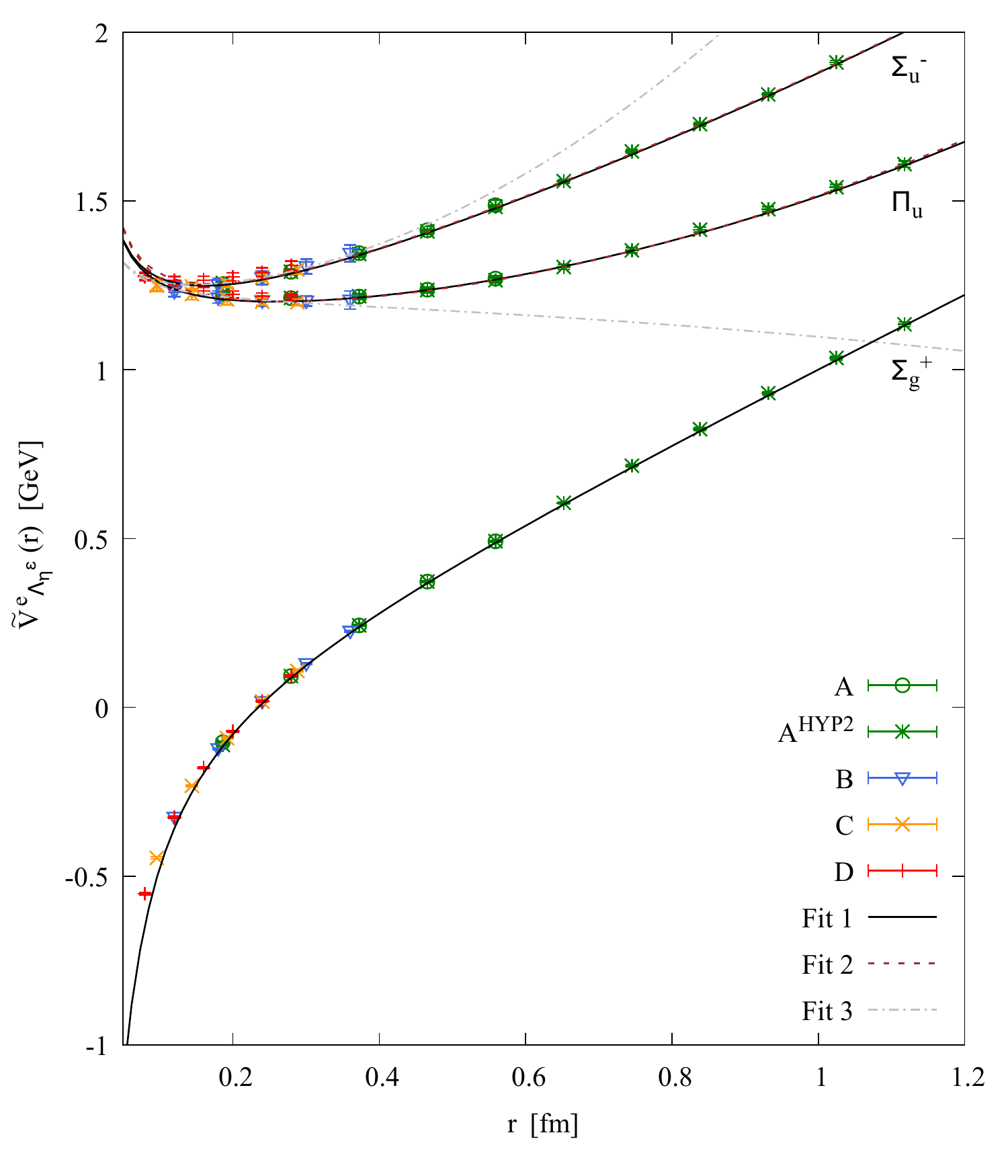}
		\caption{Lattice data points $\tilde{V}_{\Lambda_\eta^\epsilon}^e(r)$ in $\text{GeV}$ and corresponding parametrizations \eqref{eq:Cornellpotential}, \eqref{eq:parametrizationHybrid},\eqref{eq:extendedparametrizationHybrid_Piu} and \eqref{eq:extendedparametrizationHybrid_Sigmau} as functions of the quark-antiquark separation $r$ in $\text{fm}$. 
			The colors green, blue, yellow and red indicate different lattice spacings $a = 0.093 \, \text{fm}$, $a = 0.060 \, \text{fm}$, $a = 0.048 \, \text{fm}$ and $a = 0.040 \, \text{fm}$.}
		\label{fig:parametrizedlatticedata1}
	\end{figure}

	The resulting fit parameters are collected in Table~\ref{tab:fitparameter}. In particular, we obtain \\ $\alpha = 0.289(2) = 0.0571(4) \, \text{GeV} \, \text{fm}$ and $\sigma = 1.064(4) \, \text{GeV}/\text{fm}$ in reasonable agreement with results from the literature \cite{Koma:2007jq}.
	These fit parameters allow to define data points
	\begin{equation}\label{eq:def_V_Sigmagplus_tilde}
		 \tilde{V}_{\Sigma_g^+}^e(r) = V_{\Sigma_g^+}^e(r) - C^e - \Delta V^{\text{lat},e}_{\Sigma_g^+}(r) ,
	\end{equation}
	where	the self-energies and the lattice discretization errors at tree-level are subtracted. These data points are collected in Table~\ref{tab:latticedata_ABCD_tilde_all} and plotted in Figure~\ref{fig:parametrizedlatticedata1}. They are consistently parametrized by $V_{\Sigma_g^+}(r)$ for $0.2 \, \text{fm} \leq r$ as demonstrated in the same figure.

	\begin{table}\centering
		\begin {tabular}{cccr}%
\toprule $\alpha \, [\text {GeV}\, \text {fm}]$&$\sigma \, [\text {GeV}/\text {fm}]$&$\alpha ^{\prime }\, [\text {GeV}\, \text {fm}]$&$\chi ^2_{\text {red}}$\\\midrule %
\pgfmathprintnumber [fixed,fixed zerofill,precision=4]{5.7138953e-2}(\pgfmathprintnumber [fixed,fixed zerofill,precision=0]{3.8348709e0})&\pgfmathprintnumber [fixed,fixed zerofill,precision=3]{1.06373746e0}(\pgfmathprintnumber [fixed,fixed zerofill,precision=0]{3.5628387e0})&\pgfmathprintnumber [fixed,fixed zerofill,precision=4]{7.3546918e-2}(\pgfmathprintnumber [fixed,fixed zerofill,precision=0]{2.3365845e1})&\pgfutilensuremath {0.7}\\\bottomrule %
\end {tabular}%
\vspace{0.3cm}
		\begin {tabular}{cccccccr}%
\toprule &$A_1\, [\text {GeV}\, \text {fm}]$&$A_2\, [\text {GeV}]$&$A_3\, [\text {GeV}\, \text {fm}^2]$&$B_1\, [\text {GeV}\, \text {fm}^2]$&$B_2\, [\text {fm}^{-1}]$&$B_3\, [\text {fm}^{-2}]$&$\chi ^2_{\text {red}}$\\\midrule %
Fit 1&\pgfmathprintnumber [fixed,fixed zerofill,precision=4]{1.24450223e-2}(\pgfmathprintnumber [fixed,fixed zerofill,precision=0]{8.9867752e0})&\pgfmathprintnumber [fixed,fixed zerofill,precision=3]{1.13481705e0}(\pgfmathprintnumber [fixed,precision=0]{7.9253708e0})&\pgfmathprintnumber [fixed,fixed zerofill,precision=3]{3.7202162e-1}(\pgfmathprintnumber [fixed,fixed zerofill,precision=0]{6.670813e0})&\pgfmathprintnumber [fixed,fixed zerofill,precision=2]{1.56367802e0}(\pgfmathprintnumber [fixed,fixed zerofill,precision=0]{1.5459961e1})&\pgfmathprintnumber [fixed,fixed zerofill,precision=1]{1.22110382600e+00}(\pgfmathprintnumber [fixed,fixed zerofill,precision=0]{3.1523804e0})&\pgfmathprintnumber [fixed,fixed zerofill,precision=1]{2.06503171462e+00}(\pgfmathprintnumber [fixed,fixed zerofill,precision=0]{1.5268661e0})&\pgfutilensuremath {1.2}\\%
Fit 2&\pgfmathprintnumber [fixed,fixed zerofill,precision=4]{1.47425299e-2}(\pgfmathprintnumber [fixed,fixed zerofill,precision=0]{1.7531021e1})&\pgfmathprintnumber [fixed,fixed zerofill,precision=3]{1.12584137e0}(\pgfmathprintnumber [fixed,precision=0]{1.1136963e1})&\pgfmathprintnumber [fixed,fixed zerofill,precision=3]{3.8141216e-1}(\pgfmathprintnumber [fixed,fixed zerofill,precision=0]{7.3682541e0})&\pgfmathprintnumber [fixed,fixed zerofill,precision=2]{1.56909521e0}(\pgfmathprintnumber [fixed,fixed zerofill,precision=0]{1.6915451e1})&\pgfmathprintnumber [fixed,fixed zerofill,precision=1]{1.01963465005e+00}(\pgfmathprintnumber [fixed,fixed zerofill,precision=0]{3.5564102e0})&\pgfmathprintnumber [fixed,fixed zerofill,precision=1]{2.27928296345e+00}(\pgfmathprintnumber [fixed,fixed zerofill,precision=0]{1.6193405e0})&\pgfutilensuremath {0.8}\\%
Fit 3&\pgfmathprintnumber [fixed,fixed zerofill,precision=4]{6.5270772e-3}(\pgfmathprintnumber [fixed,fixed zerofill,precision=0]{1.6047684e1})&\pgfmathprintnumber [fixed,fixed zerofill,precision=3]{1.1901189e0}(\pgfmathprintnumber [fixed,precision=0]{1.3920792e1})&\pgfmathprintnumber [fixed,fixed zerofill,precision=3]{-9.231802e-2}(\pgfmathprintnumber [fixed,fixed zerofill,precision=0]{9.1465912e1})&\pgfmathprintnumber [fixed,fixed zerofill,precision=2]{1.15475719e0}(\pgfmathprintnumber [fixed,fixed zerofill,precision=0]{3.7553528e0})&-&-&\pgfutilensuremath {0.5}\\\bottomrule %
\end {tabular}%
\vspace{0.3cm}
		\begin {tabular}{ccrlrlrll}%
\toprule \multicolumn {2}{c}{}& \multicolumn {2}{c}{Fit 1}& \multicolumn {2}{c}{Fit 2}& \multicolumn {2}{c}{Fit 3} \\ensemble&$C^e\, [\text {GeV}]$&$ A'^e_{2,\Pi _u}$&$ A'^e_{2,\Sigma _u^-}$&$ A'^e_{2,\Pi _u}$&$ A'^e_{2,\Sigma _u^-}$&$ A'^e_{2,\Pi _u}$&$ A'^e_{2,\Sigma _u^-}$&$[\text {GeV}/\text {fm}^2]$\\\midrule %
$A$&\pgfmathprintnumber [fixed,fixed zerofill,precision=3]{1.39807323e0}(\pgfmathprintnumber [fixed,fixed zerofill,precision=0]{2.45887e0})&\pgfmathprintnumber [fixed,fixed zerofill,precision=1]{3.1180925e0}(\pgfmathprintnumber [fixed,fixed zerofill,precision=0]{7.1589249e0})&\pgfmathprintnumber [fixed,fixed zerofill,precision=1]{6.6771326e0}(\pgfmathprintnumber [fixed,fixed zerofill,precision=0]{7.7946335e0})&\pgfmathprintnumber [fixed,fixed zerofill,precision=1]{3.0002983e0}(\pgfmathprintnumber [fixed,fixed zerofill,precision=0]{8.6701279e0})&\pgfmathprintnumber [fixed,fixed zerofill,precision=1]{6.5213602e0}(\pgfmathprintnumber [fixed,fixed zerofill,precision=0]{8.6701279e0})&\pgfmathprintnumber [fixed,fixed zerofill,precision=1]{3.3996489e0}(\pgfmathprintnumber [fixed,fixed zerofill,precision=0]{7.9592896e0})&\pgfmathprintnumber [fixed,fixed zerofill,precision=1]{5.7392262e0}(\pgfmathprintnumber [fixed,fixed zerofill,precision=0]{9.4745346e0})&\\%
$B$&\pgfmathprintnumber [fixed,fixed zerofill,precision=3]{2.0585054e0}(\pgfmathprintnumber [fixed,fixed zerofill,precision=0]{2.4840805e0})&&&&&&&\\%
$C$&\pgfmathprintnumber [fixed,fixed zerofill,precision=3]{2.4717497e0}(\pgfmathprintnumber [fixed,fixed zerofill,precision=0]{2.459578e0})&&&&&&&\\%
$D$&\pgfmathprintnumber [fixed,fixed zerofill,precision=3]{2.8624449e0}(\pgfmathprintnumber [fixed,fixed zerofill,precision=0]{2.4601913e0})&&&&&&&\\%
$A^{\text {HYP2}}$&\pgfmathprintnumber [fixed,fixed zerofill,precision=3]{3.4015456e-1}(\pgfmathprintnumber [fixed,fixed zerofill,precision=0]{2.3365845e0})&\pgfmathprintnumber [fixed,fixed zerofill,precision=1]{9.7952913e-1}(\pgfmathprintnumber [fixed,fixed zerofill,precision=0]{7.4682892e0})&\pgfmathprintnumber [fixed,fixed zerofill,precision=1]{5.0160509e0}(\pgfmathprintnumber [fixed,fixed zerofill,precision=0]{5.2628952e0})&\pgfmathprintnumber [fixed,fixed zerofill,precision=1]{8.7551794e-1}(\pgfmathprintnumber [fixed,fixed zerofill,precision=0]{8.626915e0})&\pgfmathprintnumber [fixed,fixed zerofill,precision=1]{4.6666304e0}(\pgfmathprintnumber [fixed,fixed zerofill,precision=0]{8.626915e0})&\pgfmathprintnumber [fixed,fixed zerofill,precision=1]{1.63131503e0}(\pgfmathprintnumber [fixed,fixed zerofill,precision=0]{7.1271606e0})&\pgfmathprintnumber [fixed,fixed zerofill,precision=1]{4.4006145e0}(\pgfmathprintnumber [fixed,fixed zerofill,precision=0]{5.7613174e0})&\\\bottomrule %
\end {tabular}%

		\caption{Resulting fit parameters.
			Fit~1 and Fit~2 correspond to the parametrizations \eqref{eq:extendedparametrizationHybrid_Piu} and  \eqref{eq:extendedparametrizationHybrid_Sigmau} and fit ranges $2a \leq r$ and $3a \leq r$, respectively. 
			Fit~3 corresponds to the parametrization \eqref{eq:parametrizationHybrid} and fit range $2a \leq r \leq 0.3 \, \text{fm}$, where $A_{3,\Pi_u}=A_3$ and $A_{3,\Sigma_u^-}=A_3 + B_1$.
	 	}
 		\label{tab:fitparameter}
	\end{table}

	Our parametrization of the $\Pi_u$ and $\Sigma_u^-$ hybrid static potentials is based on the pNRQCD prediction for small separations $r \ll 1/\Lambda_{\text{QCD}}$,
	\begin{equation}
		V^{\text{pNRQCD}}_{\text{hybrid}} (r) = V_o (r) + \Lambda_H + \order{r^2}.
	\end{equation}
	(see Refs.\ \cite{Brambilla:1999xf,Berwein:2015vca}).
	pNRQCD hybrid static energies are given through the perturbative octet potential $V_o(r)$ and a non-perturbative constant $\Lambda_H$ at leading order in a multipole expansion.
	The next term in such a multipole expansion is proportional to $r^2$.
	At leading order in perturbation theory, $V_o(r) \propto 1/r$.
	
	Simple fit functions consistent with this pNRQCD prediction are
	\begin{equation}
		\label{eq:parametrizationHybrid}
		V_{\Lambda_\eta^\epsilon}(r) = \frac{A_1}{r}+ A_2 + A_{3,\Lambda_\eta^\epsilon} r^2 ,
	\end{equation}
	where the parameters $A_1$ and $A_2$ are the same both for the $\Pi_u$ and the $\Sigma_u^-$ hybrid static potential, while the coefficients in front of the $r^2$ terms, $A_{3,\Pi_u}$ and $A_{3,\Sigma_u^-}$, are independent.
	As in our preceding work \cite{Capitani:2018rox}, we found that Eq.\ \eqref{eq:parametrizationHybrid} is suited to parametrize the $\Pi_u$ potential in the $r$ range, where lattice data points are available, but not suited to parametrize the $\Sigma_u^-$ potential.
	Because of that we use extended fit functions already proposed in Ref.\ \cite{Capitani:2018rox},
	\begin{align}
		\label{eq:extendedparametrizationHybrid_Piu}
		V_{\Pi_u}(r) &= \frac{A_1}{r} + A_2 + A_3 r^2 \\
		\label{eq:extendedparametrizationHybrid_Sigmau}
		V_{\Sigma_u^-}(r) &= \frac{A_1}{r} + A_2 + A_3 r^2 + \frac{B_1 r^2}{1 + B_2 r + B_3 r^2} ,
	\end{align}
	which reduce to Eq.\ \eqref{eq:parametrizationHybrid} in the limit of small separations.

	Note that, in principle, all fit parameters depend on the lattice spacing $a$.
	In practice, however, only $A_2$ seems to have a sizable $a$ dependence, as indicated by a small ensemble dependent additive offset particularly prominent at large $a$.
	We, thus, include the leading order lattice discretization error for $A_2$, which is proportional to $a^2$. 
	It can be different for the $\Pi_u$ and the $\Sigma_u^-$ hybrid static potential and when using HYP2 smeared temporal links or not, i.e.\ is represented by terms $A'^e_{2,\Lambda_\eta^\epsilon} a^2$ with $A'^A_{2,\Pi_u} = A'^B_{2,\Pi_u} = A'^C_{2,\Pi_u} = A'^D_{2,\Pi_u}$ and $A'^A_{2,\Sigma_u^-} = A'^B_{2,\Sigma_u^-} = A'^C_{2,\Sigma_u^-} = A'^D_{2,\Sigma_u^-}$.

	As previously for the ordinary static potential, we also include in the fit functions the constants $C^e$ containing the self-energies. Moreover, we include a term reflecting discretization errors at tree level,
	\begin{equation}\label{eq:latticeartifacts_hybrid}
		\Delta V^{\text{lat},e}_\text{hybrid}(r) = 	-\frac{1}{8}\Delta V^{\text{lat},e}_{\Sigma_g^+}(r) = -\frac{\alpha'}{8} \bigg(\frac{1}{r} - \frac{G^e(r/a)}{a}\bigg) ,
	\end{equation}
	where the prefactor $-1 / 8$ relative to Eq.\ (\ref{eq:latticeartifacts}) is motivated by leading order perturbation theory.
  	In summary, this amounts to a $10$-parameter uncorrelated $\chi^2$ minimizing fit of
	\begin{align}
		\label{eq:extendedparametrizationHybrid_Piu_A2a}
		V^{\text{fit},e}_{\Pi_u}(r) &= V_{\Pi_u}(r) + C^e +\Delta V^{\text{lat},e}_\text{hybrid} (r) + A'^e_{2,\Pi_u} a^2\\
		\label{eq:extendedparametrizationHybrid_Sigmau_A2a}
		V^{\text{fit},e}_{\Sigma_u^-}(r) &= V_{\Sigma_u^-}(r) + C^e+\Delta V^{\text{lat},e}_\text{hybrid} (r) + A'^e_{2,\Sigma_u^-} a^2
	\end{align}
	to the lattice data points $V^e_{\Pi_u}(r)$ and $V^e_{\Sigma_u^-}(r)$ of all five ensembles.

	In Table~\ref{tab:fitparameter} we compare results obtained with two fit ranges, $2a \leq r$ (Fit~1) and $3a \leq r$ (Fit~2). 
	In analogy to Eq.~\eqref{eq:def_V_Sigmagplus_tilde} we define data points
	\begin{align}\label{eq:def_V_hybrid_tilde}
		\tilde{V}^e_{\Lambda_\eta^\epsilon}(r) = V^e_{\Lambda_\eta^\epsilon}(r) - C^e - \Delta V^{\text{lat},e}_\text{hybrid} (r) - A'^e_{2,\Lambda_\eta^\epsilon} a^2 ,
	\end{align}
	where the self-energy as well as lattice discretization errors at tree-level and proportional to $a^2$ in the difference to the ordinary static potential are removed. These data points are collected in Table~\ref{tab:latticedata_ABCD_tilde_all} and plotted in Figure~\ref{fig:parametrizedlatticedata1} together with the parametrizations \eqref{eq:extendedparametrizationHybrid_Piu} and \eqref{eq:extendedparametrizationHybrid_Sigmau}.
	For larger separations, $r \gtapprox 0.2 \, \text{fm}$, the parametrizations corresponding to $2a \leq r$ and to $3a \leq r$ are quite similar. 
	For separations $r \ltapprox 0.15 \, \text{fm}$, however, there are clear deviations, which signal the importance of computing data points at small $r$. 
	This is also reflected by the difference in the results for the coefficient $A_1$ of the repulsive $1/r$ term, $A_1 = 0.0124(9)\,\text{GeV} \, \text{fm}$ versus \\ $A_1 = 0.0147(18)\,\text{GeV} \, \text{fm}$ for $2a \leq r$ and $3a \leq r$, respectively.
	Since the corresponding reduced $\chi^2$ (listed in Table~\ref{tab:fitparameter}) indicate that both fits are of reasonable quality, we consider the parametrization obtained by taking into account a larger number of data points (i.e.\ Fit~1 with with $2a \leq r$) to be superior and recommend to use this parametrization in future applications, e.g.\ Born-Oppenheimer predictions of heavy hybrid meson masses. 
	
	To study hybrid static potentials at small separations in even more detail, we performed an additional fit, where we fixed $B_2 = B_3 = 0$. 
	The fit ansatz is then equivalent to Eq.\ (\ref{eq:parametrizationHybrid}), when identifying $A_{3,\Pi_u}$ in Eq.\ \eqref{eq:parametrizationHybrid} with $A_3$ in Eq.\ \eqref{eq:extendedparametrizationHybrid_Piu} and $A_{3,\Sigma_u^-}$ in Eq.\ \eqref{eq:parametrizationHybrid} with $A_3 + B_1$ in Eq.\ \eqref{eq:extendedparametrizationHybrid_Sigmau}.
	Since the fit ansatz is then restricted to the perturbative prediction valid for small $r$, we use a reduced fit range, $2a \leq r \leq 0.3 \, \text{fm}$. 
	The fit is of reasonable quality and as before the fit results are collected in Table~\ref{tab:fitparameter} and the corresponding parametrization is shown in Figure~\ref{fig:parametrizedlatticedata1}. 
	The coefficient of the repulsive $1/r$ term is now significantly smaller, $A_1 = 0.0065(16) \,\text{GeV} \, \text{fm}$. 
	
	In summary, our lattice data points, both for the $\Pi_u$ and the $\Sigma_u^-$ hybrid static potential clearly show a repulsive behavior at small separations, as predicted perturbatively in pNRQCD. 
	We performed various fits with fit functions guided by these perturbative expansions, which are proportional to $1/r$ at small separations. 
	We find the coefficient $A_1$ of the $1/r$ term in the region $0.005 \, \text{GeV} \, \text{fm} \ldots 0.017 \, \text{GeV} \, \text{fm}$. 
	A more precise determination of a parametrization of the repulsive region of hybrid static potentials will require further data points at even smaller separations and possibly refined fit functions with additional terms contributing to the small-$r$ behavior. 
	
	Finally we compare to existing work, where the $1/r$ repulsion of the $\Pi_u$ and the $\Sigma_u^-$ hybrid static potentials was also quantified.	
	In Ref.\ \cite{Vairo:2009tn} the lattice data from Ref.\ \cite{Juge:2002br} for the $\Pi_u$ hybrid static potential was parametrized with a fit function similar to Eq.\ \eqref{eq:parametrizationHybrid} with $A_1=0.022 \, \text{GeV} \, \text{fm}$, which is larger than our results for $A_1$ from simultaneous fits to the $\Pi_u$ and $\Sigma_u^-$ potentials.
	Ref.\ \cite{Oncala:2017hop} follows the prediction from  perturbation theory at leading order in $\alpha_s$ and fixes the $1/r$-coefficient to $\alpha/8$, where $\alpha$ is obtained from a fit similar to Eq.~\eqref{eq:Cornellpotential} to lattice data from Ref.\ \cite{Juge:2002br} in the range $0.2\,\text{fm} \le r \le 2.4\,\text{fm}$.
	The resulting $1/r$-coefficient for the hybrid potentials is $0.012 \, \text{GeV} \, \text{fm}$, which agrees with our fit results for the parameter $A_1$ for Fit~1 and Fit~2.
	In Ref.\ \cite{Berwein:2015vca} hybrid static potential lattice data from Refs.\ \cite{Bali:2003jq,Juge:2002br} is parametrized consistently at small separations $0.08 \,\text{fm} \le r < 0.25\,\text{fm}$ by functions similar to Eq.~\eqref{eq:parametrizationHybrid}. There, the $1/r$-coefficient is not a fit parameter, but fixed to $\approx 0.01 \, \text{GeV} \, \text{fm}$ by the perturbative octet potential calculated in the Renormalon Subtracted scheme up to order $\alpha^3_s$. This value for the $1/r$-coefficient is in the ballpark of our fit results for $A_1$.


	\subsection{Prediction of masses of heavy hybrid mesons}

		In the following we estimate masses of $\bar c c$ and $\bar b b$ hybrid mesons following the same Born-Oppenheimer approach as in our previous work \cite{Capitani:2018rox}, this time, however, using the refined and more accurate parametrizations (\ref{eq:Cornellpotential}), (\ref{eq:extendedparametrizationHybrid_Piu}) and (\ref{eq:extendedparametrizationHybrid_Sigmau}) with parameters corresponding to Fit~1 (see Table~\ref{tab:fitparameter}). Our goal is to quantify the impact of our new lattice data (results for ensembles $A$, $B$, $C$ and $D$), which cover smaller separations as well as several smaller lattice spacings than our previous data from ensemble $A^\text{HYP2}$.

		We solve the radial Schr\"odinger equation
		\begin{eqnarray}\label{radialSchroedingerequation}
			 \bigg(-\frac{1}{2 \mu} \frac{d^2}{dr^2} + \frac{L (L+1) - 2 \Lambda^2 + J_{\Lambda_{\eta}^{\epsilon}} (J_{\Lambda_{\eta}^{\epsilon}}+1)}{2 \mu r^2} + V_{\Lambda_{\eta}^{\epsilon}}(r)\bigg) u_{\Lambda_{\eta}^{\epsilon};L,n}(r) \ \ = \ \ E_{\Lambda_{\eta}^{\epsilon};L,n} u_{\Lambda_{\eta}^{\epsilon};L,n}(r) ,
		\end{eqnarray}
		where $\mu = m_{\bar{Q}} m_Q /  (m_{\bar{Q}} + m_Q)$ is the reduced mass of the heavy $\bar{Q} Q$ pair, $L$ is the orbital angular momentum and $V_{\Lambda_{\eta}^{\epsilon}}(r)$ is one of our parametrizations \eqref{eq:Cornellpotential}, \eqref{eq:extendedparametrizationHybrid_Piu} or \eqref{eq:extendedparametrizationHybrid_Sigmau} with parameters listed as Fit~1 in Table~\ref{tab:fitparameter}.
		We use $ m_c = 1628 \, \textrm{MeV}$ and $ m_b = 4977 \, \textrm{MeV}$ from quark models \cite{PhysRevD.32.189}.
		$J_{\Sigma_g^+} = 0$ and $J_{\Lambda_{\eta}^{\epsilon}} = 1$ for $\Lambda_{\eta}^{\epsilon} = \Pi_u , \Sigma_u^-$ following Ref.\ \cite{Braaten:2014qka}.
		For further details on the numerical solution of the radial Schrödinger equation and the interpretation of the resulting energies in terms of hybrid meson multiplets we refer to Ref.\ \cite{Capitani:2018rox}.

		In Table~\ref{tab:hybridmesonmasses} we provide our updated results for heavy hybrid meson masses, which are defined according to
		\begin{eqnarray}
			\label{EQN864} m_{\Lambda_{\eta}^{\epsilon};L,n} \ \ = \ \ E_{\Lambda_{\eta}^{\epsilon};L,n} - E_{\Lambda_{\eta}^{\epsilon}=\Sigma_g^+;n=1,L=0} + \overline{m} ,
		\end{eqnarray}
		where $\overline{m}$ is the spin averaged mass of the lightest quarkonium from experiments, either \\ $\overline{m} = (m_{\eta_c(1S),\textrm{exp}} + 3 m_{J/\Psi(1S),\textrm{exp}}) / 4 = 3.069(1) \, \textrm{GeV}$ or \\ $\overline{m} = (m_{\eta_b(1S),\textrm{exp}} + 3 m_{\Upsilon(1S),\textrm{exp}}) / 4 = 9.445(1) \, \textrm{GeV}$ \cite{ParticleDataGroup:2018ovx}.
		In particular the masses obtained with $V_{\Sigma_u^-}(r)$ are around $55 \, \text{MeV}$ lower for $\bar c c$ and $35 \, \text{MeV}$ lower for $\bar b b$ compared to our previous results from Ref.\ \cite{Capitani:2018rox}. The masses related to $V_{\Pi_u}(r)$ are around $20 \, \text{MeV}$ lower for $\bar c c$ and almost unchanged for $\bar b b$. 
		These discrepancies are similar to our newly introduced term $A'^e_{2,\Lambda_\eta^\epsilon} a^2$ evaluated for $e = A^\text{HYP2}$, which is $43(4) \, \text{MeV}$ for $\Lambda_\eta^\epsilon = \Sigma_u^-$ and $9(6) \, \text{MeV}$ for $\Lambda_\eta^\epsilon = \Pi_u$. 
		The term $A'^e_{2,\Lambda_\eta^\epsilon} a^2$ represents lattice discretization errors, and can only be determined, when static potential lattice data is available for several lattice spacings. 
		This demonstrates that the lattice data and the corresponding parametrizations provided in this work constitute an important step towards higher precision in Born-Oppenheimer predictions of heavy hybrid meson masses.
		The remaining discrepancies seem to be mostly related to the coefficient $\alpha$ in the parametrization (\ref{eq:Cornellpotential}) of $V_{\Sigma_g^+}(r)$, for which we quoted $\alpha = 0.0518(5) \, \text{GeV} \, \text{fm}$ in Ref.\ \cite{Capitani:2018rox} and which we updated to $\alpha = 0.0571(4) \, \text{GeV} \, \text{fm}$ in this work. 
		This change in $\alpha$ might also be a consequence of our careful identification and removal of lattice discretization errors, this time related to the tree level improvement represented by the term $\Delta V^{\text{lat},e}_{\Sigma_g^+}(r)$ defined in Eq.\ (\ref{eq:latticeartifacts}). 

		\begin{table}[htb]
		  \centering
			\begin {tabular}{ccccc}%
\toprule $\Lambda _{\eta }^{\epsilon }$&$L$&$n$&$ m_{\Lambda _{\eta }^{\epsilon };L,n}$ in GeV for $Q\bar {Q}=c\bar {c}$&$ m_{\Lambda _{\eta }^{\epsilon };L,n}$ in GeV for $Q\bar {Q}=b\bar {b}$\\\midrule %
\multirow {4}{*}{$\Pi _u$}&\pgfutilensuremath {1}&\pgfutilensuremath {1}&\pgfmathprintnumber [fixed,fixed zerofill,precision=3]{4.17480779808e+00} (\pgfmathprintnumber [fixed,fixed zerofill,precision=0]{6.4225647e0})&\pgfmathprintnumber [fixed,fixed zerofill,precision=3]{1.06817077740e+01} (\pgfmathprintnumber [fixed,fixed zerofill,precision=0]{5.8351471e0})\\%
&\pgfutilensuremath {1}&\pgfutilensuremath {2}&\pgfmathprintnumber [fixed,fixed zerofill,precision=3]{4.55025967268e+00} (\pgfmathprintnumber [fixed,fixed zerofill,precision=0]{8.2191605e0})&\pgfmathprintnumber [fixed,fixed zerofill,precision=3]{1.08951615067e+01} (\pgfmathprintnumber [fixed,fixed zerofill,precision=0]{6.3687958e0})\\%
&\pgfutilensuremath {2}&\pgfutilensuremath {1}&\pgfmathprintnumber [fixed,fixed zerofill,precision=3]{4.35978574775e+00} (\pgfmathprintnumber [fixed,fixed zerofill,precision=0]{7.247496e0})&\pgfmathprintnumber [fixed,fixed zerofill,precision=3]{1.07848702889e+01} (\pgfmathprintnumber [fixed,fixed zerofill,precision=0]{6.1674011e0})\\%
&\pgfutilensuremath {3}&\pgfutilensuremath {1}&\pgfmathprintnumber [fixed,fixed zerofill,precision=3]{4.54631424600e+00} (\pgfmathprintnumber [fixed,fixed zerofill,precision=0]{8.272818e0})&\pgfmathprintnumber [fixed,fixed zerofill,precision=3]{1.08900111094e+01} (\pgfmathprintnumber [fixed,fixed zerofill,precision=0]{6.5307663e0})\\%
\midrule \multirow {6}{*}{$\Sigma _u^-$}&\pgfutilensuremath {0}&\pgfutilensuremath {1}&\pgfmathprintnumber [fixed,fixed zerofill,precision=3]{4.43876299172e+00} (\pgfmathprintnumber [fixed,fixed zerofill,precision=0]{4.6976563e0})&\pgfmathprintnumber [fixed,fixed zerofill,precision=3]{1.08756776024e+01} (\pgfmathprintnumber [fixed,fixed zerofill,precision=0]{4.6654236e0})\\%
&\pgfutilensuremath {0}&\pgfutilensuremath {2}&\pgfmathprintnumber [fixed,fixed zerofill,precision=3]{4.87797542892e+00} (\pgfmathprintnumber [fixed,fixed zerofill,precision=0]{5.2903122e0})&\pgfmathprintnumber [fixed,fixed zerofill,precision=3]{1.11527040924e+01} (\pgfmathprintnumber [fixed,fixed zerofill,precision=0]{4.7145782e0})\\%
&\pgfutilensuremath {1}&\pgfutilensuremath {1}&\pgfmathprintnumber [fixed,fixed zerofill,precision=3]{4.57387837220e+00} (\pgfmathprintnumber [fixed,fixed zerofill,precision=0]{4.7841034e0})&\pgfmathprintnumber [fixed,fixed zerofill,precision=3]{1.09600564133e+01} (\pgfmathprintnumber [fixed,fixed zerofill,precision=0]{4.6769455e0})\\%
&\pgfutilensuremath {1}&\pgfutilensuremath {2}&\pgfmathprintnumber [fixed,fixed zerofill,precision=3]{5.00119383032e+00} (\pgfmathprintnumber [fixed,fixed zerofill,precision=0]{5.532248e0})&\pgfmathprintnumber [fixed,fixed zerofill,precision=3]{1.12282958837e+01} (\pgfmathprintnumber [fixed,fixed zerofill,precision=0]{4.7968216e0})\\%
&\pgfutilensuremath {2}&\pgfutilensuremath {1}&\pgfmathprintnumber [fixed,fixed zerofill,precision=3]{4.76192813710e+00} (\pgfmathprintnumber [fixed,fixed zerofill,precision=0]{4.9717331e0})&\pgfmathprintnumber [fixed,fixed zerofill,precision=3]{1.10782663249e+01} (\pgfmathprintnumber [fixed,fixed zerofill,precision=0]{4.6986984e0})\\%
&\pgfutilensuremath {3}&\pgfutilensuremath {1}&\pgfmathprintnumber [fixed,fixed zerofill,precision=3]{4.96373623041e+00} (\pgfmathprintnumber [fixed,fixed zerofill,precision=0]{5.2721573e0})&\pgfmathprintnumber [fixed,fixed zerofill,precision=3]{1.12051305985e+01} (\pgfmathprintnumber [fixed,fixed zerofill,precision=0]{4.7513321e0})\\\bottomrule %
\end {tabular}%

			\caption{Predictions for heavy hybrid meson masses.}
			\label{tab:hybridmesonmasses}
		\end{table}

		We note that our prediction of heavy hybrid meson masses within the Born-Oppenheimer approximation is based on several limiting assumptions (see the discussion in Section~$6.2$ in Ref.\ \cite{Capitani:2018rox}), e.g.\ the single-channel approximation, where mixing between static potentials is excluded, and the neglect of effects due to the heavy quark spins.
		More sophisticated coupled channel Schrödinger equations were derived and used for Born-Oppenheimer predictions in Refs.\ \cite{Berwein:2015vca,Oncala:2017hop}.
		Moreover, in Refs.\ \cite{Brambilla:2018pyn,Brambilla:2019jfi} first steps were taken to include corrections from the heavy quark spins.
		These more advanced approaches also require lattice field theory results for the ordinary static potential and the $\Pi_u$ and $\Sigma_u^-$ hybrid static potentials. 
		However, the corresponding predictions of heavy hybrid meson masses are based on lattice field theory results obtained at significantly larger lattice spacing than our smallest lattice spacing and corresponding parametrizations trustworthy only at larger quark-antiquark separations and presumably suffering from sizable lattice discretization errors. 
		It would be interesting to repeat the Born-Oppenheimer computations from Refs.\ \cite{Berwein:2015vca,Oncala:2017hop,Brambilla:2018pyn,Brambilla:2019jfi} with the lattice field theory results for static potentials from this work provided in Table~\ref{tab:latticedata_ABCD_tilde_all}.

		Finally we note that a precision determination of heavy hybrid meson masses in a Born-Oppenheimer framework also requires precise knowledge of static potentials for separations even smaller than $0.08 \, \text{fm}$, for which lattice computations were carried out in this work. 
		In this small-$r$ region higher order perturbation theory might be more suited than lattice QCD. 
		In Ref.\ \cite{Karbstein:2018mzo} the combination of NNNLO perturbation theory and lattice QCD is discussed for the $\Sigma_g^+$ potential. 
		It would be worthwhile to advance in the same direction for hybrid static potentials.

	\section{Excluding systematic errors}\label{sec:systematicerrors}

	\subsection{Topological freezing}
	
		In the continuum, gauge field configurations can be classified according to their integer topological charge $Q$. The corresponding topological sectors are separated by barriers of infinite action.
		
		Topological freezing refers to the problem that a Monte Carlo simulation of a lattice gauge theory is trapped in one of the topological sectors, either during a significant part or the whole simulation.
		Clearly, gauge link configurations generated in such a simulation do not form a representative set distributed according to $e^{-S}$.
		Topological freezing is expected to appear, when using small lattice spacings $a \approx 0.05 \, \text{fm}$ \cite{Schaefer:2010hu}. 
		It becomes increasingly more problematic, when approaching the continuum, i.e.\ when further decreasing $a$.
		If a simulation is fully trapped in a topological sector, observables exhibit specific finite volume corrections proportional to powers of $1/V$ ($V$ denotes the spacetime volume) \cite{Brower:2003yx,Aoki:2007ka,Dromard:2014ela,Bietenholz:2016ymo} in addition to finite volume corrections not related to topological freezing, which are discussed in Section~\ref{sec:finitevolume}.
		
\begin{figure}
		  \begin{center}
			\begin{minipage}{0.7\linewidth}
				\includegraphics[width=\linewidth]{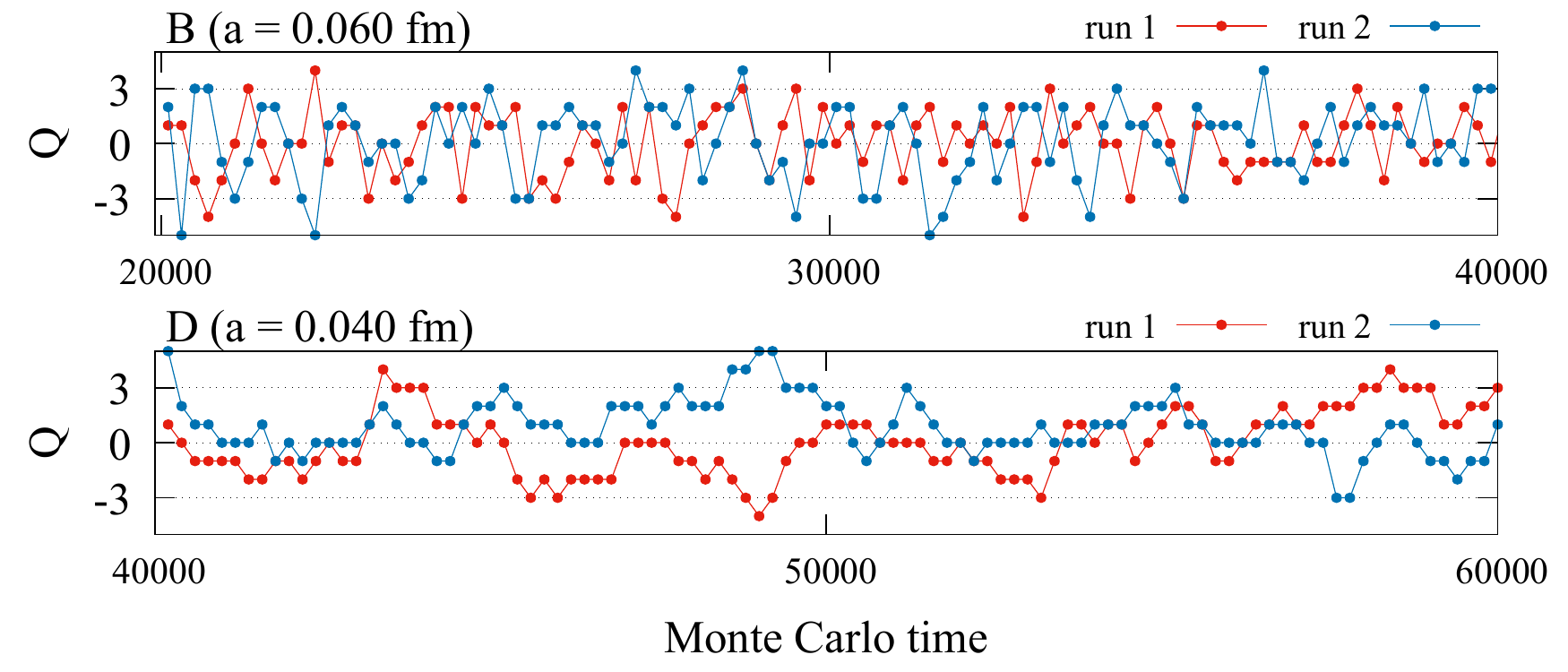}
				\subcaption{}
				\label{fig:topcharge}
			\end{minipage} \\
			\begin{minipage}{0.7\linewidth}
				\includegraphics[width=\linewidth,page=3]{plots/Plot_toplogy}
				\subcaption{}
				\label{fig:topchargehistogram}
			\end{minipage}
			\end{center}
			\caption{(a)~Monte Carlo histories of the topological charge for ensemble $B$ ($a = 0.06 \, \text{fm}$) and $D$ ($a = 0.04 \, \text{fm}$) for two independent simulation runs. (b)~Normalized and symmetrized histograms reflecting the topological charge distribution for ensemble $B$ and ensemble $D$.}
		\end{figure}

		Since our lattice spacings are as small as $0.04 \, \text{fm}$, we consider it important and neccessary, to check and compare the Monte Carlo histories of the topological charge for all our simulations.
		We use a field strength definition of the topological charge on the lattice (for a discussion and comparison of various definitions see Refs.\ \cite{Cichy:2014qta,Alexandrou:2017hqw}),
		\begin{equation}
			Q = a^4\sum_x q(x)
		\end{equation}
		with the clover-leaf discretization of the topological charge density
		\begin{eqnarray}
			& & q(x) = \frac{1}{32\pi^2} \sum_{\mu,\nu,\sigma,\rho=0}^{3}\epsilon_{\mu \nu \rho \sigma} \Tr \left(C^{\textrm{clov}}_{\mu \nu}(x) C^{\textrm{clov}}_{\rho \sigma} (x)\right) \\
			& & C^{\textrm{clov}}_{\mu \nu}(x) = \frac{1}{4} \Im \Big(P_{\mu \nu}(x)+ P_{\nu -\mu}(x)+P_{-\mu -\nu}(x)+P_{-\nu \mu}(x)\Big).
		\end{eqnarray}
		To eliminate UV-fluctuations, which do not contribute to the topological charge, but might cause strong distortions of the corresponding lattice results, a smoothing procedure needs to be applied to the gauge links.
		We use 4-dimensional APE-smearing, similar to the 3-dimensional APE-smearing for the static potential operators, with $\alpha_{\text{APE}}=0.3$.
		The number of smearing steps is chosen individually for each lattice spacing. 
		We stop smearing, as soon as $Q$ is stable for several smearing steps for the majority of gauge link configurations.
		We computed the topological charge on all gauge link configurations of the four ensembles $A$, $B$, $C$ and $D$ given in Table~\ref{tab:latticesetups4}.
		In Figure~\ref{fig:topcharge} we show exemplarily the Monte Carlo histories of the topological charge for a subset of gauge link configurations for ensemble $B$ ($a = 0.06 \, \text{fm}$) and $D$ ($a = 0.04 \, \text{fm}$).
		At $a = 0.06 \, \text{fm}$ the topological charge changes frequently and topological freezing is clearly not a problem.
		At $a = 0.04 \, \text{fm}$ the autocorrelation time of $Q$ is much longer, consistent with the expectation from Ref.\ \cite{Schaefer:2010hu}. 
		However, there are still sufficiently many changes, such that our statistical error analysis, based on four independent simulation runs and a suitable binning (see Appendix~\ref{APP012}), should provide realistic uncertainties for the static potentials.
		
		In Figure~\ref{fig:topchargehistogram} we show normalized and symmetrized histograms reflecting the topological charge distribution for ensembles $B$ and $D$. 
		Both are consistent with Gaussian distributions, as expected at finite, large spacetime volume.
		From their squared widths, $\expval{Q^2}$, we obtain estimates of the related topological susceptibilities via $\chi_{\text{top}}= \expval{Q^2}/V$, which are in reasonable agreement with results from the literature \cite{Athenodorou:2021qvs}. 
		This is another indication that our computations of static potentials do not suffer from the problem of topological freezing.
		
	\subsection{Finite volume corrections}\label{sec:finitevolume}
		All static potential results discussed in Section~\ref{sec:latticeresults} and Section~\ref{sec:parametrization} were obtained from simulations with periodic spatial volume $L^3  \approx (1.2 \, \text{fm})^3 $. 
		Since this is a rather small volume, it is important to check that finite volume corrections to these results are negligible.
		
		One source of finite volume corrections are virtual glueballs traveling around the far side of the periodic spacetime volume. 
		They cause a negative shift of energy levels, which is proportional to $\exp(-m_{0^{++}}L)$ at asymptotically large $L$\cite{Luscher:1985dn} ($m_{0^{++}}$ denotes the mass of the lightest glueball).
		We observe a small negative shift for the ordinary static potential for $L \ll 1.0 \, \text{fm}$, which could be related to such glueball interactions.
		Another type of finite volume correction will appear, when the (infinite volume) wave function of a state has a larger extent than the finite spacetime volume of the lattice. 
		Then this wave function is necessarily squeezed, which will lead to a positive shift of the corresponding energy level \cite{Fukugita:1992jj}.
		For the $\Pi_u$ and $\Sigma_u^-$ hybrid static potentials we found sizable positive shifts for $L \ll 1.0 \, \text{fm}$. 
		Since their wave functions cover a significantly larger region than the ordinary static potential \cite{Muller:2019joq}, these positive shifts are also consistent with expectation.
		
\begin{figure}\centering
			\includegraphics[width=0.5\linewidth,page=2]{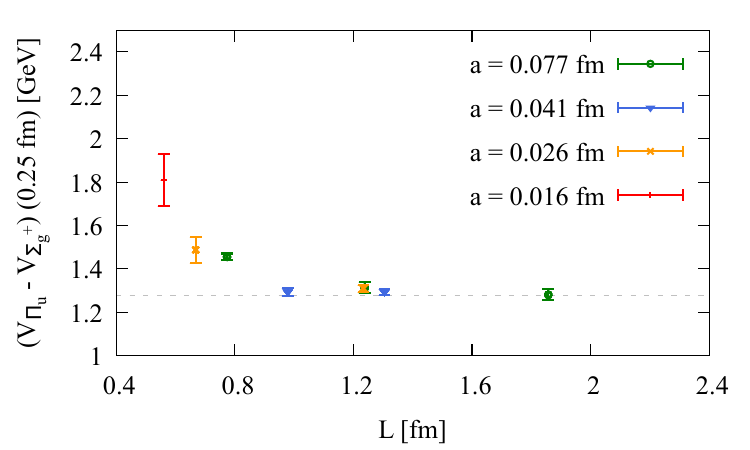}
			\caption{$(V_{\Pi_u}(0.25\,\text{fm}) - V_{\Sigma_g^+}(0.25\,\text{fm}))$ as function of the spatial lattice extent $L$ for gauge group $\text{SU(2)}$. Data points were obtained at several lattice spacings, but discretization errors were estimated to be smaller than statistical errors.}
			\label{fig:FVE_potentialdifference_SU2}
		\end{figure}

		In Figure~\ref{fig:FVE_potentialdifference_SU2} we show the difference between the $\Pi_u$ hybrid static potential and the ordinary static potential, $V_{\Pi_u} - V_{\Sigma_g^+}$, at fixed quark-antiquark separation $r=0.25 \, \text{fm}$ as function of the spatial lattice extent $L$ for gauge group $\text{SU(2)}$. 
		This difference is consistent with a constant, i.e.\ $L$-independent, for $L \gtapprox 1.0 \, \text{fm}$. 
		For smaller $L$, however, the difference increases, which is consistent with the previously discussed expectation of a squeezed wave function for the $\Pi_u$ hybrid static potential. 
		
		Additionally, for gauge group $\text{SU(3)}$ we compared results for the $\Sigma_g^+$, the $\Pi_u$ and the $\Sigma_u^-$ static potential from ensemble $A$ to results from an analogous computation with twice the spatial lattice extent, i.e.\ $L = 2.4 \, \text{fm}$. 
		We did not find statistically significant differences.
		
		In summary, the investigations and checks discussed in this subsection strongly indicate that finite volume corrections at our preferred spacetime volume $L^3 \times T \approx (1.2 \, \text{fm})^3 \times 2.4 \, \text{fm}$ are small compared to current statistical errors and, thus, can be neglected.

	\subsection{Glueball decay}\label{sec:glueballdecay}
\begin{figure}\centering
			\includegraphics[width=\linewidth]{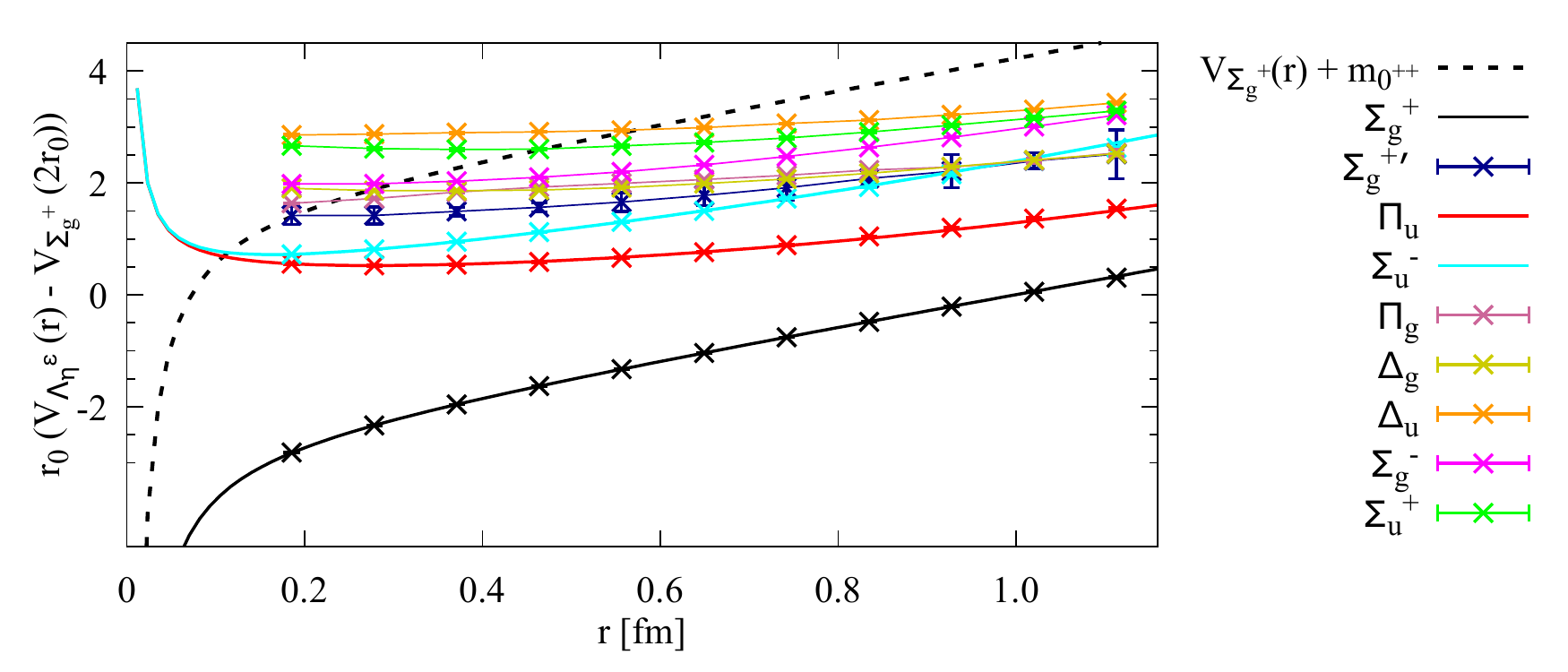}
			\caption{Threshold energy for decays of hybrid flux tubes into the $\Sigma_g^+$ ground state and a $0^{++}$ glueball (dashed line) and hybrid static potentials for various quantum numbers $\Lambda_\eta^\epsilon$. Static potentials are taken from Ref.\ \cite{Capitani:2018rox}, the $0^{++}$ glueball mass from Ref.\ \cite{Morningstar:1999rf}.}
				\label{fig:threshold}
		\end{figure}

		At sufficiently small $r$, the energy difference between a hybrid static potential and the ordinary static potential is large enough such that the $\Lambda_\eta^\epsilon$ hybrid flux tube can decay into a glueball and the $\Sigma_g^+$ groundstate.
		The threshold energy for a decay into the lightest glueball with quantum numbers $J^{PC}=0^{++}$ and mass $m_{0^{++}}=1.73(5) \,\text{GeV}$ \cite{Morningstar:1999rf} is shown as dashed line in Figure~\ref{fig:threshold} together with lattice results for hybrid static potentials from Ref.\ \cite{Capitani:2018rox}.
		The critical separations $r^{\Lambda_\eta^\epsilon}_{\text{crit}}$, where the dashed line intersects the $\Lambda_\eta^\epsilon$ hybrid static potentials, are listed in Table \ref{tab:rcrit}. For $r \leq r^{\Lambda_\eta^\epsilon}_{\text{crit}}$ decays to a $0^{++}$ glueball are energetically allowed.

		\begin{table}[hbt]\centering
			\renewcommand{\arraystretch}{1.5}
			\begin{tabular}{c|cccccc|cc}\hline
				$\Lambda_{\eta}^{\epsilon}$ & $\Pi_u$    & $\Pi_g$    & $\Delta_g$ & $\Delta_u$ & ${{\Sigma}_g^+}^{\prime}$ & $\Sigma_u^+$ & $\Sigma_u^-$ & $\Sigma_g^-$\\ \hline
				$r^{\Lambda_\eta^\epsilon}_{\text{crit}}\,  [\text{fm}]$           & $0.11$ & $0.23 $ & $0.28 $ & $0.58 $ &$0.19 $  &$0.46 $  &$0.11 $  & $0.3 $ \\\hline
			\end{tabular}
			\caption{Maximal separation $r_{\text{crit}}^{\Lambda_\eta^\epsilon}$, where a decay of a $\Lambda_\eta^\epsilon$ hybrid flux tube into the $\Sigma_g^+$ ground state and a $0^{++}$ glueball is energetically possible. For $\Sigma_u^-$ and $\Sigma_g^-$ such decays are excluded, because of quantum numbers.
			}
			\label{tab:rcrit}
			\renewcommand{\arraystretch}{1.0}
		\end{table}
							
		However, such decays might be excluded, because of quantum numbers. 
		A comprehensive and general derivation of selection rules for both hybrids and tetraquarks can be found in Ref.\ \cite{Braaten:2014ita}.
		Here we focus on hybrid static potentials with quantum numbers $\Lambda^\epsilon_\eta$ and discuss, whether decays to the $\Sigma_g^+$ groundstate and a $J^{PC} = 0^{++}$ glueball are possible.
		Since $J = 0$ for the considered glueball, also $J_z = 0$.
		Thus, the $z$-component of the orbital angular momentum of the glueball must be $L_z = \Lambda$ (as stated in Section~\ref{sec:theory_hybridstaticpotentials}, the static quark and antiquark are separated along the $z$ axis).
		The quantum number $\eta$ does not protect a hybrid flux tube, because the distribution of the glueball in $z$-direction can be symmetric or antisymmetric.
		There is, however, a constraint due to the quantum number $\epsilon$.
		The $0^{++}$ glueball is symmetric with respect to $\mathcal{P}_x$.
		Its orbital angular momentum wave function is also symmetric with respect to $\mathcal{P}_x$ for $L_z = \Lambda = \Sigma = 0$.
		For $L_z = \Lambda > 0$ there are two independent possibilities for the wave function, one of them symmetric, the other antisymmetric.
		From this one can conclude that a $0^{++}$ glueball decay is not possible for $\Sigma_u^-$ and for $\Sigma_g^-$, while it is allowed for all other hybrid flux tubes.
		
		Decays into heavier glueballs with quantum numbers $J^{PC}$ different from $0^{++}$ (some of them are antisymmetric with respect to $\mathcal{P}_x$) are energetically only allowed for separations much smaller than those listed in Table~\ref{tab:rcrit}.
		Thus, they are not relevant in the context of our work.
		
		In Section~\ref{sec:latticeresults} and Section~\ref{sec:parametrization} we presented and used lattice results for separations as small as $r \approx 0.08 \, \text{fm}$. Since $r_{\text{crit}}^{\Pi_u} = 0.11 \, \text{fm}$, results for the $\Pi_u$ hybrid static potential below that separation might be contaminated by a ``$\Sigma_g^+$ + glueball'' scattering state.
		However, we observe the expected upward curvature for the $\Pi_u$ hybrid static potential (see Figure~\ref{fig:parametrizedlatticedata1}).
		Moreover, the $\Pi_u$ und $\Sigma_u^-$ hybrid static potentials approach each other for small $r$, consistent with the expected degeneracy in the limit $r \rightarrow 0$.
		Thus, we conclude that a possible contamination of our results for the $\Pi_u$ hybrid static potential is negligible compared to statistical errors.

	\section{Summary and outlook}

We computed the ordinary static potential and the $\Pi_u$ and $\Sigma_u^-$ hybrid static potentials in $\text{SU(3)}$ lattice gauge theory at four different lattice spacings, where the smallest lattice spacing, $a = 0.04 \, \text{fm}$, is roughly half the size of lattice spacings previously used in similar computations. 
Lattice discretization errors, which were found to be sizable in the bare lattice data points, were studied in detail. 
We removed a large part of these discretization errors by using both perturbative tree-level improvement and a suitable simultaneous fit to the bare lattice data points from all our ensembles  to identify the dominant $a^2$ contribution to the discretization errors. 
Using the same fit we were also able to subtract the $a$-dependent unphysical self-energy. 
For future reference these improved lattice data points are collected in Table~\ref{tab:latticedata_ABCD_tilde_all}. 
Moreover, we investigated possibly existent systematic errors related to topological freezing, due to the finite spatial volume and because of glueball decays in detail and provided evidence that these errors are negligible compared to statistical errors.

We also provide parametrizations of the $\Sigma_g^+$, $\Pi_u$ and $\Sigma_u^-$ static potentials, which can e.g.\ be used for Born-Oppenheimer predictions of heavy hybrid meson masses. 
The Born-Oppenheimer approach in the context of heavy hybrid mesons received considerable interest in the past couple of years, with many improvements and refinements, e.g.\ the derivation of coupled channel Schr\"odinger equations, which take into account mixing between different sectors \cite{Berwein:2015vca,Oncala:2017hop}, or the inclusion of effects due to the heavy quark spins \cite{Brambilla:2018pyn,Brambilla:2019jfi}. 
These papers, however, use lattice data \cite{Juge:1997nc,Juge:1997ir,Juge:1999aw,Juge:2002br,Bali:2003jq,Juge:2003ge} generated around two decades ago at much coarser lattice spacing and partly without any dedicated investigation or removal of discretization errors. 
Thus, it would be an interesting and important step towards higher precision to combine the refined Born-Oppenheimer approaches from Refs.\ \cite{Berwein:2015vca,Oncala:2017hop,Brambilla:2018pyn,Brambilla:2019jfi} with the lattice data points or the parametrizations presented in this work.

Since we performed computations at very small lattice spacings, we were able to reliably access quark-antiquark separations as small as $r = 0.08 \, \text{fm}$. 
This, in turn, allowed to convincingly show the upward curvature at small $r$ of the $\Pi_u$ and $\Sigma_u^-$ hybrid static potentials predicted by perturbation theory, i.e.\ their repulsive nature at small quark-antiquark separations. 
An interesting future direction with the aim to improve the precision of Born-Oppenheimer predictions even further could be to match higher order perturbation theory and the lattice results presented in this work. 
For the ordinary static potential a possible method using next-to-next-to-next-to-leading order perturbation theory was discussed in Ref.\ \cite{Karbstein:2018mzo} and an approach based on leading order perturbation theory for hybrid static potentials can be found in Ref.\ \cite{Berwein:2015vca}.

\section*{Acknowledgements}

We thank Christian Reisinger for providing his multilevel code.
We acknowledge interesting and useful discussions with Eric Braaten, Nora Brambilla, Francesco Knechtli, Colin Morningstar, Lasse M\"uller, Christian Reisinger and Joan Soto.

M.W.\ acknowledges support by the Heisenberg Programme of the Deutsche Forschungsgemeinschaft (DFG, German Research Foundation) -- project number 399217702.

Calculations on the GOETHE-HLR and on the on the FUCHS-CSC high-performance computers of the Frankfurt University were conducted for this research. We would like to thank HPC-Hessen, funded by the State Ministry of Higher Education, Research and the Arts, for programming advice.

	\appendix
	
	\section{Optimization of operators} \label{Appendix:optimization}
		
		The operators $S$ appearing in Eq.\ (\ref{eq:a}) were optimized in Ref.\ \cite{Capitani:2018rox} at lattice spacing $a=0.093 \, \text{fm}$ with respect to their generated ground state overlaps. 
		To retain this optimization also for smaller values of the lattice spacing, we adjust the operator extents in $x$- and $y$-direction in units of the lattice spacing, $E_x$ and $E_y$, such that the operator extents in physical units, $E_x a$ and $E_y a$, are almost independent of $a$.
		Moreover, we select $N_\text{APE}$, the number of APE-smearing steps, individually for each lattice spacing, while keeping $\alpha_\text{APE} = 0.5$ constant (see e.g.\ Ref.\ \cite{Jansen:2008si} for detailed equations). 
		We do this in such a way that the effective potentials of the $\Pi_u$ and the $\Sigma_u^-$ hybrid static potentials at temporal separation $t/a = 2$ are small. 
		This amounts to increasing $N_\text{APE}$ for decreasing $a$. Our preferred values for $N_\text{APE}$ both for gauge group $\text{SU(2)}$ and $\text{SU(3)}$ are listed in Table~\ref{TAB500}.
		
		\begin{table}[htb]
			\begin{center}
				\begin{tabular}{ccccc}
					\toprule
					$a$ in $\text{fm}$ & $0.078$ & $0.041$ & $0.026$ \\
					\midrule
					$N_\text{APE}$ for $\text{SU(2)}$ & $30$ & $100$ & $200$ \\
					\bottomrule %
				\end{tabular}
				
				\vspace{0.3cm}
				\begin{tabular}{ccccc}
					\toprule
					$a$ in $\text{fm}$ & $0.093$ & $0.060$ & $0.048$ & $0.040$ \\
					\midrule
					$N_\text{APE}$ for $\text{SU(3)}$ & $20$ & $50$ & $75$ & $100$ \\
					\bottomrule %
				\end{tabular}
			\end{center}
			
			\caption{\label{TAB500}Smearing parameter $N_\text{APE}$ for various lattice spacings for gauge group $\text{SU(2)}$ and $\text{SU(3)}$.}
		\end{table}



	\section{\label{APP012}Error analysis}
	
		To eliminate correlations in Monte Carlo time, we combine consecutively generated gauge link configurations, which are used for the computation of static potentials, to $N^e$ bins.
		
		For the data points $V^e_{\Lambda_\eta^\epsilon}(r)$ (see Section~\ref{sec:latticeresults}) statistical errors are determined via a standard jackknife analysis, i.e.\ from $N^e$ reduced jackknife samples $V^{e,\text{jackknife}}_{\Lambda_\eta^\epsilon,j}(r)$ according to
		\begin{align}
			\Delta V^e_{\Lambda_\eta^\epsilon}(r) = \bigg(\frac{N^e - 1}{N^e} \sum_{j=1}^{N^e} \Big(V^{e,\text{jackknife}}_{\Lambda_\eta^\epsilon,j}(r) - \bar{V}^e_{\Lambda_\eta^\epsilon}(r)\Big)^2\bigg)^{1/2}
		\end{align}
		($\bar{V}^e_{\Lambda_\eta^\epsilon}(r)$ denotes the result for the full sample).
		
		The fits from Section~\ref{sec:parametrization}, where data points of all five ensembles are used at the same time, can in principle also be computed via the jackknife method. 
		The number of reduced jackknife samples, however, would be rather large, $N^A \times N^B \times N^C \times N^D \times N^{A^\text{HYP2}}$, and the corresponding computational effort huge. 
		Therefore, we use for these fits and all following analyses the bootstrap method. 
		To this end we first inflate the reduced jackknife samples,
		\begin{align}
			V^e_{\Lambda_\eta^\epsilon,j}(r) = \bar{V}^e_{\Lambda_\eta^\epsilon}(r) + (N^e - 1) \Big(V^{e,\text{jackknife}}_{\Lambda_\eta^\epsilon,j}(r) - \bar{V}^e_{\Lambda_\eta^\epsilon}(r)\Big) .
		\end{align}
		A bootstrap sample is then generated by randomly selecting $N^e$ of the inflated samples $V^e_{\Lambda_\eta^\epsilon,j}(r)$ for each ensemble, where the same inflated sample may be selected more than once. 
		As usual, the bootstrap error of a quantity $O$ is then the standard deviation of the results obtained on the bootstrap samples, i.e.\
		\begin{align}
			\Delta O = \bigg(\frac{1}{K} \sum_{j=1}^K \Big(O_j - \bar{O}\Big)^2\bigg)^{1/2} .
		\end{align}
		$O_j$ denotes the result on the $j$-th bootstrap sample and $\bar{O}$ the result on the full sample, where $O$ can be $\alpha$, $\sigma$, $\alpha'$, $C^e$, $A_1$, $A_2$, etc. $K$, the number of bootstrap samples, has to be chosen sufficiently large, such that $\Delta O$ is essentially independent of $K$.
		
		For our computations we used $N^A = 320$, $ N^B = N^C = N^D = 160$, $N^{A^\text{HYP2}} = 5000$ and $K = 10000$.


	\section{Tree-level improvement}\label{Appendix:treelevelimprovement}
		
		In the continuum at tree-level of perturbation theory the ordinary static potential potential is attractive and proportional to $1/r$. 
		Its lattice counterpart for the standard Wilson plaquette gauge action and the Eichten-Hill static action is
		\begin{equation}\label{eq:definition_rimproved}
			\left(\frac{1}{r}\right)_\text{lat} = 4\pi G(r/a,0,0),
		\end{equation}
		where the Greens function 
		\begin{equation}\label{eq:latticepropagator1}
			G(\textbf{R}) = \frac{1}{(2\pi)^3} \int_{-\pi}^{\pi} \dd[3]{k} 
			\frac{\prod_{j=1}^{3} \cos(k_j R_j)}{4 \sum_{j=1}^{3} \sin[2](k_j/2)},
		\end{equation} 
		can be computed in an efficient way via a recursion relation \cite{Luscher:1995zz,Necco:2001xg}.
		
		For the HYP2 static action \cite{DellaMorte:2005nwx,Hasenfratz:2001hp,DellaMorte:2003mn}, which was used for the computations on ensemble $A^\text{HYP2}$, the numerator of the integrand differs from Eq.\ \eqref{eq:latticepropagator1} by an additional  factor,
		\begin{equation}\label{eq:latticepropagator2}
			G^\text{HYP}(\textbf{R}) = \frac{1}{(2\pi)^3} \int_{-\pi}^{\pi} \dd[3]{k} 
			\frac{\prod_{j=1}^{3} \cos(k_j R_j) \times \Big(
				1- (\alpha_1 / 6) \sum_{i=1}^{3}4\sin[2](k_i)\Omega_{i0}
				\Big)^2}{4 \sum_{j=1}^{3} \sin[2](k_j/2)},
		\end{equation} 
		where $\Omega_{\mu \nu}$ is
		\begin{align}
			\Omega_{\mu \nu} = 1+ &\alpha_2(1+\alpha_3) - \frac{\alpha_2}{4}(1+2\alpha_3)\bigg(
			\sum_{j=1}^{3}4\sin[2](p_j/2) - 4\sin[2](p_{\mu}/2) - 4\sin[2](p_{\nu}/2)\bigg) \\\nonumber
			&+ \frac{\alpha_2 \alpha_3}{4}\prod_{\eta \neq \mu,\nu}^{} 4\sin[2](p_{\mu}/2).
		\end{align}
		(see Ref.\ \cite{Hasenfratz:2001tw}).
		This integral can be solved e.g.\ by standard Monte Carlo integration techniques.
		
		To eliminate lattice discretization errors at tree-level for the ordinary static potential, we subtract
		\begin{equation}
			\Delta V^{\text{lat},e}_{\Sigma_g^+}(r) = \alpha' \bigg(\frac{1}{r} - \frac{G^e(r/a)}{a}\bigg)
		\end{equation}
		from the lattice data points (see Section~\ref{sec:parametrization}, in particular Eq.\ (\ref{eq:def_V_Sigmagplus_tilde})), where $G^e(r/a) = 4 \pi G(r/a,0,0)$ for $e=A,B,C,D$ and $G^e(r/a) = 4 \pi G^\text{HYP}(r/a,0,0)$ for $e=A^{HYP2}$. $\alpha'$ is proportional to the strong coupling and determined by a fit to the lattice data (see again Section~\ref{sec:parametrization}).
		Similarly, we subtract $\Delta V^{\text{lat},e}_\text{hybrid}(r) = -(1/8) \Delta V^{\text{lat},e}_{\Sigma_g^+}(r)$ from the lattice data points for the $\Pi_u$ and $\Sigma_u^-$ hybrid static potentials, which are repulsive and, in leading-order perturbation theory, suppressed by the factor $1/8$ relative to the ordinary static potential.
		
\begin{figure}[htb]
			\centering
			\includegraphics[width=\linewidth,page=1]{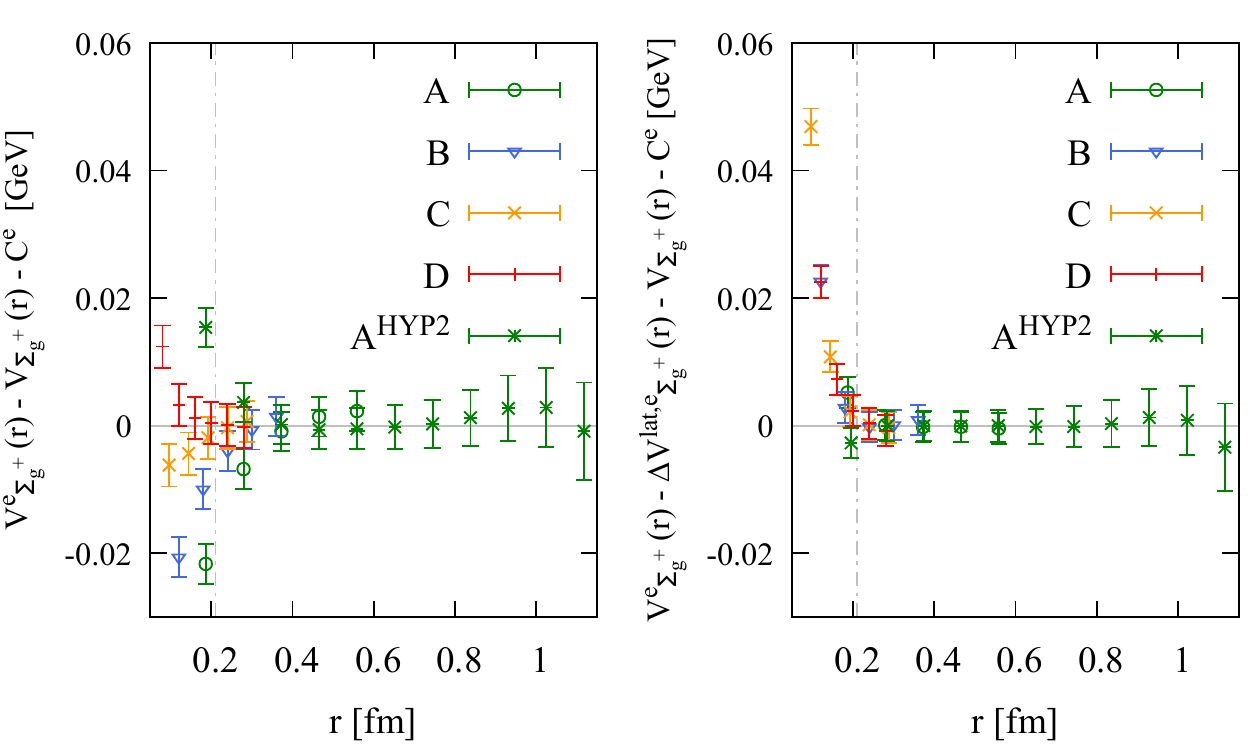}
			\caption{Comparison of unimproved (left) and improved (right) lattice data points for the $\Sigma_g^+$ static potential  from our five ensembles $A$, $B$, $C$, $D$ and $A^\text{HYP2}$.
				We subtract $V_{\Sigma_g^+}(r) + C^e$ with $V_{\Sigma_g^+}(r)$ defined in Eq.\ \eqref{eq:Cornellpotential} and parameters obtained by a fit to data points with $r \geq 0.2 \, \text{fm}$ (indicated by the vertical dashed line) as listed in Table~\ref{tab:fitparameter}.}
			\label{fig:data_mi_para}
		\end{figure}

		The benefit of applying tree-level improvement, when combining lattice field theory results obtained at different lattice spacings and with different static actions is demonstrated in Figure~\ref{fig:data_mi_para}, where we compare unimproved and improved data points for the $\Sigma_g^+$ potential from our five ensembles. 
		The two plots show that the majority of improved data points are consistent with a single curve, while unimproved data points from different ensembles exhibit strong discrepancies for $r \ltapprox 0.4 \, \text{fm}$.
		
	\section{Summary of $\text{SU(3)}$ lattice field theory results for the $\Sigma_g^+$, $\Pi_u$ and $\Sigma_u^-$ static potentials}\label{Appendix:summarylatticedata}

		In Table~\ref{tab:latticedata_ABCD_all} we list $V^e_{\Lambda_{\eta}^{\epsilon}}(r)a$, the bare lattice data points in units of the lattice spacing (see Section~\ref{sec:latticeresults}). 
		In Table~\ref{tab:latticedata_ABCD_tilde_all} we list $\tilde{V}^e_{\Lambda_\eta^\epsilon}(r)$, the lattice data points defined in Eqs.\ (\ref{eq:def_V_Sigmagplus_tilde}) and (\ref{eq:def_V_hybrid_tilde}), where the self energy as well as lattice discretization errors at tree-level and proportional to $a^2$ are removed.
		
		\renewcommand{\arraystretch}{1.1}
		\begin{table}[h]\centering
			\begin {tabular}{ccccc}%
\toprule ensemble&$r/a$&$V^e_{\Sigma _g^+ }\, a$&$V^e_{\Pi _u }\, a$&$V^e_{\Sigma _u^- }\, a$\\\midrule %
\multirow {5}{*}{$A$}&\pgfutilensuremath {2}&\pgfmathprintnumber [fixed,fixed zerofill,precision=6]{5.96753464540e-01}(\pgfmathprintnumber [fixed,fixed zerofill, set thousands separator={},precision=0]{6.2022903e1})&\pgfmathprintnumber [fixed,fixed zerofill,precision=4]{1.25273942160e+00}(\pgfmathprintnumber [fixed,fixed zerofill,precision=0]{2.7860046e1})&\pgfmathprintnumber [fixed,fixed zerofill,precision=4]{1.28130006703e+00}(\pgfmathprintnumber [fixed,fixed zerofill,precision=0]{3.5481445e1})\\%
&\pgfutilensuremath {3}&\pgfmathprintnumber [fixed,fixed zerofill,precision=6]{6.99138279880e-01}(\pgfmathprintnumber [fixed,fixed zerofill, set thousands separator={},precision=0]{1.337764e2})&\pgfmathprintnumber [fixed,fixed zerofill,precision=4]{1.24571522214e+00}(\pgfmathprintnumber [fixed,fixed zerofill,precision=0]{2.7230286e1})&\pgfmathprintnumber [fixed,fixed zerofill,precision=4]{1.29717575365e+00}(\pgfmathprintnumber [fixed,fixed zerofill,precision=0]{4.0602814e1})\\%
&\pgfutilensuremath {4}&\pgfmathprintnumber [fixed,fixed zerofill,precision=6]{7.72857385181e-01}(\pgfmathprintnumber [fixed,fixed zerofill, set thousands separator={},precision=0]{2.3807602e2})&\pgfmathprintnumber [fixed,fixed zerofill,precision=4]{1.24757643899e+00}(\pgfmathprintnumber [fixed,fixed zerofill,precision=0]{2.714743e1})&\pgfmathprintnumber [fixed,fixed zerofill,precision=4]{1.32279099215e+00}(\pgfmathprintnumber [fixed,fixed zerofill,precision=0]{4.8121078e1})\\%
&\pgfutilensuremath {5}&\pgfmathprintnumber [fixed,fixed zerofill,precision=6]{8.35108535148e-01}(\pgfmathprintnumber [fixed,fixed zerofill, set thousands separator={},precision=0]{3.8657776e2})&\pgfmathprintnumber [fixed,fixed zerofill,precision=4]{1.25704137431e+00}(\pgfmathprintnumber [fixed,fixed zerofill,precision=0]{2.8035461e1})&\pgfmathprintnumber [fixed,fixed zerofill,precision=4]{1.35470284138e+00}(\pgfmathprintnumber [fixed,fixed zerofill,precision=0]{5.7694443e1})\\%
&\pgfutilensuremath {6}&\pgfmathprintnumber [fixed,fixed zerofill,precision=6]{8.91806901858e-01}(\pgfmathprintnumber [fixed,fixed zerofill, set thousands separator={},precision=0]{6.013176e2})&\pgfmathprintnumber [fixed,fixed zerofill,precision=4]{1.27260756729e+00}(\pgfmathprintnumber [fixed,fixed zerofill,precision=0]{3.0128906e1})&\pgfmathprintnumber [fixed,fixed zerofill,precision=4]{1.38979716224e+00}(\pgfmathprintnumber [fixed,fixed zerofill,precision=0]{7.1012482e1})\\%
\midrule \multirow {5}{*}{$B$}&\pgfutilensuremath {2}&\pgfmathprintnumber [fixed,fixed zerofill,precision=6]{5.12961248683e-01}(\pgfmathprintnumber [fixed,fixed zerofill, set thousands separator={},precision=0]{1.8509491e1})&\pgfmathprintnumber [fixed,fixed zerofill,precision=4]{1.00665132121e+00}(\pgfmathprintnumber [fixed,fixed zerofill,precision=0]{4.5820908e1})&\pgfmathprintnumber [fixed,fixed zerofill,precision=4]{1.01503113243e+00}(\pgfmathprintnumber [fixed,fixed zerofill,precision=0]{5.1831482e1})\\%
&\pgfutilensuremath {3}&\pgfmathprintnumber [fixed,fixed zerofill,precision=6]{5.84399784947e-01}(\pgfmathprintnumber [fixed,fixed zerofill, set thousands separator={},precision=0]{4.0265976e1})&\pgfmathprintnumber [fixed,fixed zerofill,precision=4]{9.99630623896e-01}(\pgfmathprintnumber [fixed,fixed zerofill,precision=0]{4.4582565e1})&\pgfmathprintnumber [fixed,fixed zerofill,precision=4]{1.01551970053e+00}(\pgfmathprintnumber [fixed,fixed zerofill,precision=0]{5.5546616e1})\\%
&\pgfutilensuremath {4}&\pgfmathprintnumber [fixed,fixed zerofill,precision=6]{6.29977104787e-01}(\pgfmathprintnumber [fixed,fixed zerofill, set thousands separator={},precision=0]{7.2747604e1})&\pgfmathprintnumber [fixed,fixed zerofill,precision=4]{9.96597156825e-01}(\pgfmathprintnumber [fixed,fixed zerofill,precision=0]{4.3652237e1})&\pgfmathprintnumber [fixed,fixed zerofill,precision=4]{1.02118266769e+00}(\pgfmathprintnumber [fixed,fixed zerofill,precision=0]{6.0600769e1})\\%
&\pgfutilensuremath {5}&\pgfmathprintnumber [fixed,fixed zerofill,precision=6]{6.64971902348e-01}(\pgfmathprintnumber [fixed,fixed zerofill, set thousands separator={},precision=0]{1.1749146e2})&\pgfmathprintnumber [fixed,fixed zerofill,precision=4]{9.96346075954e-01}(\pgfmathprintnumber [fixed,fixed zerofill,precision=0]{4.3329727e1})&\pgfmathprintnumber [fixed,fixed zerofill,precision=4]{1.03147390841e+00}(\pgfmathprintnumber [fixed,fixed zerofill,precision=0]{6.7132141e1})\\%
&\pgfutilensuremath {6}&\pgfmathprintnumber [fixed,fixed zerofill,precision=6]{6.94703147205e-01}(\pgfmathprintnumber [fixed,fixed zerofill, set thousands separator={},precision=0]{1.7536957e2})&\pgfmathprintnumber [fixed,fixed zerofill,precision=4]{9.97268330856e-01}(\pgfmathprintnumber [fixed,fixed zerofill,precision=0]{8.2046371e1})&\pgfmathprintnumber [fixed,fixed zerofill,precision=4]{1.04323778601e+00}(\pgfmathprintnumber [fixed,fixed zerofill,precision=0]{7.559552e1})\\%
\midrule \multirow {5}{*}{$C$}&\pgfutilensuremath {2}&\pgfmathprintnumber [fixed,fixed zerofill,precision=6]{4.78535202336e-01}(\pgfmathprintnumber [fixed,fixed zerofill, set thousands separator={},precision=0]{1.1023026e1})&\pgfmathprintnumber [fixed,fixed zerofill,precision=4]{9.09112808622e-01}(\pgfmathprintnumber [fixed,fixed zerofill,precision=0]{4.0372147e1})&\pgfmathprintnumber [fixed,fixed zerofill,precision=4]{9.11993086978e-01}(\pgfmathprintnumber [fixed,fixed zerofill,precision=0]{4.2616913e1})\\%
&\pgfutilensuremath {3}&\pgfmathprintnumber [fixed,fixed zerofill,precision=6]{5.40197862603e-01}(\pgfmathprintnumber [fixed,fixed zerofill, set thousands separator={},precision=0]{2.5664474e1})&\pgfmathprintnumber [fixed,fixed zerofill,precision=4]{9.01852464561e-01}(\pgfmathprintnumber [fixed,fixed zerofill,precision=0]{3.8885208e1})&\pgfmathprintnumber [fixed,fixed zerofill,precision=4]{9.09471658430e-01}(\pgfmathprintnumber [fixed,fixed zerofill,precision=0]{2.4727798e1})\\%
&\pgfutilensuremath {4}&\pgfmathprintnumber [fixed,fixed zerofill,precision=6]{5.77592308012e-01}(\pgfmathprintnumber [fixed,fixed zerofill, set thousands separator={},precision=0]{4.7029129e1})&\pgfmathprintnumber [fixed,fixed zerofill,precision=4]{8.97631715500e-01}(\pgfmathprintnumber [fixed,fixed zerofill,precision=0]{3.7150696e1})&\pgfmathprintnumber [fixed,fixed zerofill,precision=4]{9.11022093136e-01}(\pgfmathprintnumber [fixed,fixed zerofill,precision=0]{2.544461e1})\\%
&\pgfutilensuremath {5}&\pgfmathprintnumber [fixed,fixed zerofill,precision=6]{6.04987989260e-01}(\pgfmathprintnumber [fixed,fixed zerofill, set thousands separator={},precision=0]{7.5524002e1})&\pgfmathprintnumber [fixed,fixed zerofill,precision=4]{8.95863269458e-01}(\pgfmathprintnumber [fixed,fixed zerofill,precision=0]{2.0869202e1})&\pgfmathprintnumber [fixed,fixed zerofill,precision=4]{9.15430537702e-01}(\pgfmathprintnumber [fixed,fixed zerofill,precision=0]{2.6398987e1})\\%
&\pgfutilensuremath {6}&\pgfmathprintnumber [fixed,fixed zerofill,precision=6]{6.27359860785e-01}(\pgfmathprintnumber [fixed,fixed zerofill, set thousands separator={},precision=0]{1.1058838e2})&\pgfmathprintnumber [fixed,fixed zerofill,precision=4]{8.95419732901e-01}(\pgfmathprintnumber [fixed,fixed zerofill,precision=0]{2.0134949e1})&\pgfmathprintnumber [fixed,fixed zerofill,precision=4]{9.21019383127e-01}(\pgfmathprintnumber [fixed,fixed zerofill,precision=0]{2.7640182e1})\\%
\midrule \multirow {6}{*}{$D$}&\pgfutilensuremath {2}&\pgfmathprintnumber [fixed,fixed zerofill,precision=6]{4.53884371348e-01}(\pgfmathprintnumber [fixed,fixed zerofill, set thousands separator={},precision=0]{6.749939e0})&\pgfmathprintnumber [fixed,fixed zerofill,precision=4]{8.41732457946e-01}(\pgfmathprintnumber [fixed,fixed zerofill,precision=0]{2.1904938e1})&\pgfmathprintnumber [fixed,fixed zerofill,precision=4]{8.43325707003e-01}(\pgfmathprintnumber [fixed,fixed zerofill,precision=0]{2.218544e1})\\%
&\pgfutilensuremath {3}&\pgfmathprintnumber [fixed,fixed zerofill,precision=6]{5.09465694509e-01}(\pgfmathprintnumber [fixed,fixed zerofill, set thousands separator={},precision=0]{1.5500214e1})&\pgfmathprintnumber [fixed,fixed zerofill,precision=4]{8.35301010749e-01}(\pgfmathprintnumber [fixed,fixed zerofill,precision=0]{2.1301498e1})&\pgfmathprintnumber [fixed,fixed zerofill,precision=4]{8.39604474770e-01}(\pgfmathprintnumber [fixed,fixed zerofill,precision=0]{2.2599014e1})\\%
&\pgfutilensuremath {4}&\pgfmathprintnumber [fixed,fixed zerofill,precision=6]{5.42057979897e-01}(\pgfmathprintnumber [fixed,fixed zerofill, set thousands separator={},precision=0]{2.8680771e1})&\pgfmathprintnumber [fixed,fixed zerofill,precision=4]{8.31690133639e-01}(\pgfmathprintnumber [fixed,fixed zerofill,precision=0]{2.059874e1})&\pgfmathprintnumber [fixed,fixed zerofill,precision=4]{8.39290337588e-01}(\pgfmathprintnumber [fixed,fixed zerofill,precision=0]{2.3180908e1})\\%
&\pgfutilensuremath {5}&\pgfmathprintnumber [fixed,fixed zerofill,precision=6]{5.65146319899e-01}(\pgfmathprintnumber [fixed,fixed zerofill, set thousands separator={},precision=0]{4.6254837e1})&\pgfmathprintnumber [fixed,fixed zerofill,precision=4]{8.29472495333e-01}(\pgfmathprintnumber [fixed,fixed zerofill,precision=0]{1.9893295e1})&\pgfmathprintnumber [fixed,fixed zerofill,precision=4]{8.41160480766e-01}(\pgfmathprintnumber [fixed,fixed zerofill,precision=0]{2.3797119e1})\\%
&\pgfutilensuremath {6}&\pgfmathprintnumber [fixed,fixed zerofill,precision=6]{5.83421583831e-01}(\pgfmathprintnumber [fixed,fixed zerofill, set thousands separator={},precision=0]{6.7998962e1})&\pgfmathprintnumber [fixed,fixed zerofill,precision=4]{8.28183198551e-01}(\pgfmathprintnumber [fixed,fixed zerofill,precision=0]{1.9202744e1})&\pgfmathprintnumber [fixed,fixed zerofill,precision=4]{8.43844064062e-01}(\pgfmathprintnumber [fixed,fixed zerofill,precision=0]{2.4740753e1})\\%
&\pgfutilensuremath {7}&\pgfmathprintnumber [fixed,fixed zerofill,precision=6]{5.98894434494e-01}(\pgfmathprintnumber [fixed,fixed zerofill, set thousands separator={},precision=0]{9.3449402e1})&\pgfmathprintnumber [fixed,fixed zerofill,precision=4]{8.27649583893e-01}(\pgfmathprintnumber [fixed,fixed zerofill,precision=0]{1.8678925e1})&\pgfmathprintnumber [fixed,fixed zerofill,precision=4]{8.47863656038e-01}(\pgfmathprintnumber [fixed,fixed zerofill,precision=0]{2.584053e1})\\%
\midrule \multirow {11}{*}{$A^{\text {HYP}2}$}&\pgfutilensuremath {2}&\pgfmathprintnumber [fixed,fixed zerofill,precision=6]{1.16647851780e-01}(\pgfmathprintnumber [fixed,fixed zerofill, set thousands separator={},precision=0]{1.2892776e1})&\pgfmathprintnumber [fixed,fixed zerofill,precision=4]{7.42680771145e-01}(\pgfmathprintnumber [fixed,fixed zerofill,precision=0]{2.1197815e1})&\pgfmathprintnumber [fixed,fixed zerofill,precision=4]{7.73650729190e-01}(\pgfmathprintnumber [fixed,fixed zerofill,precision=0]{7.4697174e0})\\%
&\pgfutilensuremath {3}&\pgfmathprintnumber [fixed,fixed zerofill,precision=6]{2.06461605656e-01}(\pgfmathprintnumber [fixed,fixed zerofill, set thousands separator={},precision=0]{3.0555328e1})&\pgfmathprintnumber [fixed,fixed zerofill,precision=4]{7.36859963775e-01}(\pgfmathprintnumber [fixed,fixed zerofill,precision=0]{1.8317749e1})&\pgfmathprintnumber [fixed,fixed zerofill,precision=4]{7.90147325501e-01}(\pgfmathprintnumber [fixed,fixed zerofill,precision=0]{8.0234436e0})\\%
&\pgfutilensuremath {4}&\pgfmathprintnumber [fixed,fixed zerofill,precision=6]{2.75767203981e-01}(\pgfmathprintnumber [fixed,fixed zerofill, set thousands separator={},precision=0]{6.0239151e1})&\pgfmathprintnumber [fixed,fixed zerofill,precision=4]{7.39503612369e-01}(\pgfmathprintnumber [fixed,fixed zerofill,precision=0]{1.7377594e1})&\pgfmathprintnumber [fixed,fixed zerofill,precision=4]{8.15089666246e-01}(\pgfmathprintnumber [fixed,fixed zerofill,precision=0]{8.9996521e0})\\%
&\pgfutilensuremath {5}&\pgfmathprintnumber [fixed,fixed zerofill,precision=6]{3.36546011885e-01}(\pgfmathprintnumber [fixed,fixed zerofill, set thousands separator={},precision=0]{1.1376862e2})&\pgfmathprintnumber [fixed,fixed zerofill,precision=4]{7.48332676778e-01}(\pgfmathprintnumber [fixed,fixed zerofill,precision=0]{1.7832352e1})&\pgfmathprintnumber [fixed,fixed zerofill,precision=4]{8.46932729739e-01}(\pgfmathprintnumber [fixed,fixed zerofill,precision=0]{1.0090683e1})\\%
&\pgfutilensuremath {6}&\pgfmathprintnumber [fixed,fixed zerofill,precision=6]{3.92895854252e-01}(\pgfmathprintnumber [fixed,fixed zerofill, set thousands separator={},precision=0]{1.8364548e2})&\pgfmathprintnumber [fixed,fixed zerofill,precision=4]{7.62073764370e-01}(\pgfmathprintnumber [fixed,fixed zerofill,precision=0]{1.93685e1})&\pgfmathprintnumber [fixed,fixed zerofill,precision=4]{8.80936873732e-01}(\pgfmathprintnumber [fixed,fixed zerofill,precision=0]{6.2367065e0})\\%
&\pgfutilensuremath {7}&\pgfmathprintnumber [fixed,fixed zerofill,precision=6]{4.46511648869e-01}(\pgfmathprintnumber [fixed,fixed zerofill, set thousands separator={},precision=0]{2.886319e2})&\pgfmathprintnumber [fixed,fixed zerofill,precision=4]{7.80500026250e-01}(\pgfmathprintnumber [fixed,fixed zerofill,precision=0]{2.1216187e1})&\pgfmathprintnumber [fixed,fixed zerofill,precision=4]{9.17109865281e-01}(\pgfmathprintnumber [fixed,fixed zerofill,precision=0]{7.1102509e0})\\%
&\pgfutilensuremath {8}&\pgfmathprintnumber [fixed,fixed zerofill,precision=6]{4.98473981062e-01}(\pgfmathprintnumber [fixed,fixed zerofill, set thousands separator={},precision=0]{4.4642288e2})&\pgfmathprintnumber [fixed,fixed zerofill,precision=4]{8.03733112966e-01}(\pgfmathprintnumber [fixed,fixed zerofill,precision=0]{2.4010559e1})&\pgfmathprintnumber [fixed,fixed zerofill,precision=4]{9.58617291288e-01}(\pgfmathprintnumber [fixed,fixed zerofill,precision=0]{8.0878555e0})\\%
&\pgfutilensuremath {9}&\pgfmathprintnumber [fixed,fixed zerofill,precision=6]{5.49516516678e-01}(\pgfmathprintnumber [fixed,fixed zerofill, set thousands separator={},precision=0]{6.7988068e2})&\pgfmathprintnumber [fixed,fixed zerofill,precision=4]{8.32597725270e-01}(\pgfmathprintnumber [fixed,fixed zerofill,precision=0]{1.539476e1})&\pgfmathprintnumber [fixed,fixed zerofill,precision=4]{9.96619825216e-01}(\pgfmathprintnumber [fixed,fixed zerofill,precision=0]{9.3462036e0})\\%
&\pgfutilensuremath {10}&\pgfmathprintnumber [fixed,fixed zerofill,precision=6]{5.99979717284e-01}(\pgfmathprintnumber [fixed,fixed zerofill, set thousands separator={},precision=0]{1.0324173e3})&\pgfmathprintnumber [fixed,fixed zerofill,precision=4]{8.61342635906e-01}(\pgfmathprintnumber [fixed,fixed zerofill,precision=0]{1.8798767e1})&\pgfmathprintnumber [fixed,fixed zerofill,precision=4]{1.03823487890e+00}(\pgfmathprintnumber [fixed,fixed zerofill,precision=0]{1.0857193e1})\\%
&\pgfutilensuremath {11}&\pgfmathprintnumber [fixed,fixed zerofill,precision=6]{6.49218153590e-01}(\pgfmathprintnumber [fixed,fixed zerofill, set thousands separator={},precision=0]{1.5628632e3})&\pgfmathprintnumber [fixed,fixed zerofill,precision=4]{8.92006849470e-01}(\pgfmathprintnumber [fixed,fixed zerofill,precision=0]{2.287294e1})&\pgfmathprintnumber [fixed,fixed zerofill,precision=4]{1.08312879457e+00}(\pgfmathprintnumber [fixed,fixed zerofill,precision=0]{1.2658081e1})\\%
&\pgfutilensuremath {12}&\pgfmathprintnumber [fixed,fixed zerofill,precision=6]{6.96190686512e-01}(\pgfmathprintnumber [fixed,fixed zerofill, set thousands separator={},precision=0]{2.3609009e3})&\pgfmathprintnumber [fixed,fixed zerofill,precision=4]{9.24250286951e-01}(\pgfmathprintnumber [fixed,fixed zerofill,precision=0]{2.7758194e1})&\pgfmathprintnumber [fixed,fixed zerofill,precision=4]{1.12661095608e+00}(\pgfmathprintnumber [fixed,fixed zerofill,precision=0]{1.4895782e1})\\\bottomrule %
\end {tabular}%

			\caption{Bare lattice data points for the $\Sigma_g^+$, $\Pi_u$ and $\Sigma_u^-$ static potentials in units of the lattice spacing (see Section~\ref{sec:latticeresults}).}
			\label{tab:latticedata_ABCD_all}
		\end{table}
		
		\begin{table}[h]\centering
			\begin {tabular}{cccccc}%
\toprule ensemble&$r/a$&$r \, [\text {fm}]$&$\tilde {V}^e_{\Sigma _g^+ }\,[\text {GeV}]$&$\tilde {V}^e_{\Pi _u }\,[\text {GeV}]$&$\tilde {V}^e_{\Sigma _u^- }\,[\text {GeV}]$\\\midrule %
\multirow {5}{*}{$A$}&\pgfutilensuremath {2}&\pgfutilensuremath {0.1863}&\pgfmathprintnumber [fixed,fixed zerofill,precision=4]{-1.0333e-1}(\pgfmathprintnumber [fixed,fixed zerofill, set thousands separator={},precision=0]{2.3985e1})&\pgfmathprintnumber [fixed,fixed zerofill,precision=4]{1.2245e0}(\pgfmathprintnumber [fixed,fixed zerofill, set thousands separator={},precision=0]{7.258e1})&\pgfmathprintnumber [fixed,fixed zerofill,precision=4]{1.2541e0}(\pgfmathprintnumber [fixed,fixed zerofill, set thousands separator={},precision=0]{7.6411e1})\\%
&\pgfutilensuremath {3}&\pgfutilensuremath {0.2794}&\pgfmathprintnumber [fixed,fixed zerofill,precision=4]{9.2882e-2}(\pgfmathprintnumber [fixed,fixed zerofill, set thousands separator={},precision=0]{2.3683e1})&\pgfmathprintnumber [fixed,fixed zerofill,precision=4]{1.2122e0}(\pgfmathprintnumber [fixed,fixed zerofill, set thousands separator={},precision=0]{7.6128e1})&\pgfmathprintnumber [fixed,fixed zerofill,precision=4]{1.2903e0}(\pgfmathprintnumber [fixed,fixed zerofill, set thousands separator={},precision=0]{9.132e1})\\%
&\pgfutilensuremath {4}&\pgfutilensuremath {0.3726}&\pgfmathprintnumber [fixed,fixed zerofill,precision=4]{2.4283e-1}(\pgfmathprintnumber [fixed,fixed zerofill, set thousands separator={},precision=0]{2.3631e1})&\pgfmathprintnumber [fixed,fixed zerofill,precision=4]{1.2169e0}(\pgfmathprintnumber [fixed,fixed zerofill, set thousands separator={},precision=0]{7.8146e1})&\pgfmathprintnumber [fixed,fixed zerofill,precision=4]{1.3454e0}(\pgfmathprintnumber [fixed,fixed zerofill, set thousands separator={},precision=0]{1.0829e2})\\%
&\pgfutilensuremath {5}&\pgfutilensuremath {0.4657}&\pgfmathprintnumber [fixed,fixed zerofill,precision=4]{3.726e-1}(\pgfmathprintnumber [fixed,fixed zerofill, set thousands separator={},precision=0]{2.3712e1})&\pgfmathprintnumber [fixed,fixed zerofill,precision=4]{1.2372e0}(\pgfmathprintnumber [fixed,fixed zerofill, set thousands separator={},precision=0]{8.0838e1})&\pgfmathprintnumber [fixed,fixed zerofill,precision=4]{1.4131e0}(\pgfmathprintnumber [fixed,fixed zerofill, set thousands separator={},precision=0]{1.267e2})\\%
&\pgfutilensuremath {6}&\pgfutilensuremath {0.5589}&\pgfmathprintnumber [fixed,fixed zerofill,precision=4]{4.9187e-1}(\pgfmathprintnumber [fixed,fixed zerofill, set thousands separator={},precision=0]{2.46e1})&\pgfmathprintnumber [fixed,fixed zerofill,precision=4]{1.2703e0}(\pgfmathprintnumber [fixed,fixed zerofill, set thousands separator={},precision=0]{8.4294e1})&\pgfmathprintnumber [fixed,fixed zerofill,precision=4]{1.4876e0}(\pgfmathprintnumber [fixed,fixed zerofill, set thousands separator={},precision=0]{1.5398e2})\\%
\midrule \multirow {5}{*}{$B$}&\pgfutilensuremath {2}&\pgfutilensuremath {0.1201}&\pgfmathprintnumber [fixed,fixed zerofill,precision=4]{-3.254e-1}(\pgfmathprintnumber [fixed,fixed zerofill, set thousands separator={},precision=0]{2.6353e1})&\pgfmathprintnumber [fixed,fixed zerofill,precision=4]{1.23175e0}(\pgfmathprintnumber [fixed,fixed zerofill, set thousands separator={},precision=0]{1.50185e2})&\pgfmathprintnumber [fixed,fixed zerofill,precision=4]{1.24646e0}(\pgfmathprintnumber [fixed,fixed zerofill, set thousands separator={},precision=0]{1.6998e2})\\%
&\pgfutilensuremath {3}&\pgfutilensuremath {0.1801}&\pgfmathprintnumber [fixed,fixed zerofill,precision=4]{-1.22707e-1}(\pgfmathprintnumber [fixed,fixed zerofill, set thousands separator={},precision=0]{2.4301e1})&\pgfmathprintnumber [fixed,fixed zerofill,precision=4]{1.21269e0}(\pgfmathprintnumber [fixed,fixed zerofill, set thousands separator={},precision=0]{1.49284e2})&\pgfmathprintnumber [fixed,fixed zerofill,precision=4]{1.25206e0}(\pgfmathprintnumber [fixed,fixed zerofill, set thousands separator={},precision=0]{1.8262e2})\\%
&\pgfutilensuremath {4}&\pgfutilensuremath {0.2402}&\pgfmathprintnumber [fixed,fixed zerofill,precision=4]{1.7427e-2}(\pgfmathprintnumber [fixed,fixed zerofill, set thousands separator={},precision=0]{2.4069e1})&\pgfmathprintnumber [fixed,fixed zerofill,precision=4]{1.20392e0}(\pgfmathprintnumber [fixed,fixed zerofill, set thousands separator={},precision=0]{1.4722e2})&\pgfmathprintnumber [fixed,fixed zerofill,precision=4]{1.27187e0}(\pgfmathprintnumber [fixed,fixed zerofill, set thousands separator={},precision=0]{1.9878e2})\\%
&\pgfutilensuremath {5}&\pgfutilensuremath {0.3002}&\pgfmathprintnumber [fixed,fixed zerofill,precision=4]{1.29192e-1}(\pgfmathprintnumber [fixed,fixed zerofill, set thousands separator={},precision=0]{2.384e1})&\pgfmathprintnumber [fixed,fixed zerofill,precision=4]{1.2035e0}(\pgfmathprintnumber [fixed,fixed zerofill, set thousands separator={},precision=0]{1.4667e2})&\pgfmathprintnumber [fixed,fixed zerofill,precision=4]{1.30609e0}(\pgfmathprintnumber [fixed,fixed zerofill, set thousands separator={},precision=0]{2.2015e2})\\%
&\pgfutilensuremath {6}&\pgfutilensuremath {0.3603}&\pgfmathprintnumber [fixed,fixed zerofill,precision=4]{2.2559e-1}(\pgfmathprintnumber [fixed,fixed zerofill, set thousands separator={},precision=0]{2.362e1})&\pgfmathprintnumber [fixed,fixed zerofill,precision=4]{1.2067e0}(\pgfmathprintnumber [fixed,fixed zerofill, set thousands separator={},precision=0]{2.6967e2})&\pgfmathprintnumber [fixed,fixed zerofill,precision=4]{1.3449e0}(\pgfmathprintnumber [fixed,fixed zerofill, set thousands separator={},precision=0]{2.4597e2})\\%
\midrule \multirow {5}{*}{$C$}&\pgfutilensuremath {2}&\pgfutilensuremath {0.0960}&\pgfmathprintnumber [fixed,fixed zerofill,precision=4]{-4.459e-1}(\pgfmathprintnumber [fixed,fixed zerofill, set thousands separator={},precision=0]{2.8288e1})&\pgfmathprintnumber [fixed,fixed zerofill,precision=4]{1.2489e0}(\pgfmathprintnumber [fixed,fixed zerofill, set thousands separator={},precision=0]{1.68292e2})&\pgfmathprintnumber [fixed,fixed zerofill,precision=4]{1.25253e0}(\pgfmathprintnumber [fixed,fixed zerofill, set thousands separator={},precision=0]{1.7951e2})\\%
&\pgfutilensuremath {3}&\pgfutilensuremath {0.1441}&\pgfmathprintnumber [fixed,fixed zerofill,precision=4]{-2.326e-1}(\pgfmathprintnumber [fixed,fixed zerofill, set thousands separator={},precision=0]{2.4464e1})&\pgfmathprintnumber [fixed,fixed zerofill,precision=4]{1.22409e0}(\pgfmathprintnumber [fixed,fixed zerofill, set thousands separator={},precision=0]{1.64668e2})&\pgfmathprintnumber [fixed,fixed zerofill,precision=4]{1.24718e0}(\pgfmathprintnumber [fixed,fixed zerofill, set thousands separator={},precision=0]{1.08281e2})\\%
&\pgfutilensuremath {4}&\pgfutilensuremath {0.1921}&\pgfmathprintnumber [fixed,fixed zerofill,precision=4]{-9.0982e-2}(\pgfmathprintnumber [fixed,fixed zerofill, set thousands separator={},precision=0]{2.4118e1})&\pgfmathprintnumber [fixed,fixed zerofill,precision=4]{1.20824e0}(\pgfmathprintnumber [fixed,fixed zerofill, set thousands separator={},precision=0]{1.5898e2})&\pgfmathprintnumber [fixed,fixed zerofill,precision=4]{1.25505e0}(\pgfmathprintnumber [fixed,fixed zerofill, set thousands separator={},precision=0]{1.1167e2})\\%
&\pgfutilensuremath {5}&\pgfutilensuremath {0.2401}&\pgfmathprintnumber [fixed,fixed zerofill,precision=4]{1.75581e-2}(\pgfmathprintnumber [fixed,fixed zerofill, set thousands separator={},precision=0]{2.4037e1})&\pgfmathprintnumber [fixed,fixed zerofill,precision=4]{1.20148e0}(\pgfmathprintnumber [fixed,fixed zerofill, set thousands separator={},precision=0]{9.5071e1})&\pgfmathprintnumber [fixed,fixed zerofill,precision=4]{1.27367e0}(\pgfmathprintnumber [fixed,fixed zerofill, set thousands separator={},precision=0]{1.16483e2})\\%
&\pgfutilensuremath {6}&\pgfutilensuremath {0.2881}&\pgfmathprintnumber [fixed,fixed zerofill,precision=4]{1.07866e-1}(\pgfmathprintnumber [fixed,fixed zerofill, set thousands separator={},precision=0]{2.4011e1})&\pgfmathprintnumber [fixed,fixed zerofill,precision=4]{1.19986e0}(\pgfmathprintnumber [fixed,fixed zerofill, set thousands separator={},precision=0]{9.2554e1})&\pgfmathprintnumber [fixed,fixed zerofill,precision=4]{1.29683e0}(\pgfmathprintnumber [fixed,fixed zerofill, set thousands separator={},precision=0]{1.21707e2})\\%
\midrule \multirow {6}{*}{$D$}&\pgfutilensuremath {2}&\pgfutilensuremath {0.0800}&\pgfmathprintnumber [fixed,fixed zerofill,precision=4]{-5.5203e-1}(\pgfmathprintnumber [fixed,fixed zerofill, set thousands separator={},precision=0]{3.03e1})&\pgfmathprintnumber [fixed,fixed zerofill,precision=4]{1.27544e0}(\pgfmathprintnumber [fixed,fixed zerofill, set thousands separator={},precision=0]{1.07825e2})&\pgfmathprintnumber [fixed,fixed zerofill,precision=4]{1.2776e0}(\pgfmathprintnumber [fixed,fixed zerofill, set thousands separator={},precision=0]{1.1304e2})\\%
&\pgfutilensuremath {3}&\pgfutilensuremath {0.1200}&\pgfmathprintnumber [fixed,fixed zerofill,precision=4]{-3.2596e-1}(\pgfmathprintnumber [fixed,fixed zerofill, set thousands separator={},precision=0]{2.47264e1})&\pgfmathprintnumber [fixed,fixed zerofill,precision=4]{1.24973e0}(\pgfmathprintnumber [fixed,fixed zerofill, set thousands separator={},precision=0]{1.08972e2})&\pgfmathprintnumber [fixed,fixed zerofill,precision=4]{1.26526e0}(\pgfmathprintnumber [fixed,fixed zerofill, set thousands separator={},precision=0]{1.16075e2})\\%
&\pgfutilensuremath {4}&\pgfutilensuremath {0.1600}&\pgfmathprintnumber [fixed,fixed zerofill,precision=4]{-1.7964e-1}(\pgfmathprintnumber [fixed,fixed zerofill, set thousands separator={},precision=0]{2.41884e1})&\pgfmathprintnumber [fixed,fixed zerofill,precision=4]{1.23372e0}(\pgfmathprintnumber [fixed,fixed zerofill, set thousands separator={},precision=0]{1.06104e2})&\pgfmathprintnumber [fixed,fixed zerofill,precision=4]{1.26552e0}(\pgfmathprintnumber [fixed,fixed zerofill, set thousands separator={},precision=0]{1.18436e2})\\%
&\pgfutilensuremath {5}&\pgfutilensuremath {0.2000}&\pgfmathprintnumber [fixed,fixed zerofill,precision=4]{-7.0586e-2}(\pgfmathprintnumber [fixed,fixed zerofill, set thousands separator={},precision=0]{2.40903e1})&\pgfmathprintnumber [fixed,fixed zerofill,precision=4]{1.22339e0}(\pgfmathprintnumber [fixed,fixed zerofill, set thousands separator={},precision=0]{1.0383e2})&\pgfmathprintnumber [fixed,fixed zerofill,precision=4]{1.27534e0}(\pgfmathprintnumber [fixed,fixed zerofill, set thousands separator={},precision=0]{1.22266e2})\\%
&\pgfutilensuremath {6}&\pgfutilensuremath {0.2400}&\pgfmathprintnumber [fixed,fixed zerofill,precision=4]{1.76212e-2}(\pgfmathprintnumber [fixed,fixed zerofill, set thousands separator={},precision=0]{2.40375e1})&\pgfmathprintnumber [fixed,fixed zerofill,precision=4]{1.21727e0}(\pgfmathprintnumber [fixed,fixed zerofill, set thousands separator={},precision=0]{1.0123e2})&\pgfmathprintnumber [fixed,fixed zerofill,precision=4]{1.28882e0}(\pgfmathprintnumber [fixed,fixed zerofill, set thousands separator={},precision=0]{1.27408e2})\\%
&\pgfutilensuremath {7}&\pgfutilensuremath {0.2800}&\pgfmathprintnumber [fixed,fixed zerofill,precision=4]{9.3025e-2}(\pgfmathprintnumber [fixed,fixed zerofill, set thousands separator={},precision=0]{2.40286e1})&\pgfmathprintnumber [fixed,fixed zerofill,precision=4]{1.21475e0}(\pgfmathprintnumber [fixed,fixed zerofill, set thousands separator={},precision=0]{9.8671e1})&\pgfmathprintnumber [fixed,fixed zerofill,precision=4]{1.30876e0}(\pgfmathprintnumber [fixed,fixed zerofill, set thousands separator={},precision=0]{1.32635e2})\\%
\midrule \multirow {11}{*}{$A^{\text {HYP}2}$}&\pgfutilensuremath {2}&\pgfutilensuremath {0.1863}&\pgfmathprintnumber [fixed,fixed zerofill,precision=4]{-1.1122e-1}(\pgfmathprintnumber [fixed,fixed zerofill, set thousands separator={},precision=0]{2.3834e1})&\pgfmathprintnumber [fixed,fixed zerofill,precision=4]{1.2267e0}(\pgfmathprintnumber [fixed,fixed zerofill, set thousands separator={},precision=0]{5.7603e1})&\pgfmathprintnumber [fixed,fixed zerofill,precision=4]{1.2573e0}(\pgfmathprintnumber [fixed,fixed zerofill, set thousands separator={},precision=0]{5.0598e1})\\%
&\pgfutilensuremath {3}&\pgfutilensuremath {0.2794}&\pgfmathprintnumber [fixed,fixed zerofill,precision=4]{9.2789e-2}(\pgfmathprintnumber [fixed,fixed zerofill, set thousands separator={},precision=0]{2.3456e1})&\pgfmathprintnumber [fixed,fixed zerofill,precision=4]{1.2126e0}(\pgfmathprintnumber [fixed,fixed zerofill, set thousands separator={},precision=0]{5.6387e1})&\pgfmathprintnumber [fixed,fixed zerofill,precision=4]{1.2904e0}(\pgfmathprintnumber [fixed,fixed zerofill, set thousands separator={},precision=0]{5.1419e1})\\%
&\pgfutilensuremath {4}&\pgfutilensuremath {0.3726}&\pgfmathprintnumber [fixed,fixed zerofill,precision=4]{2.4301e-1}(\pgfmathprintnumber [fixed,fixed zerofill, set thousands separator={},precision=0]{2.361e1})&\pgfmathprintnumber [fixed,fixed zerofill,precision=4]{1.2178e0}(\pgfmathprintnumber [fixed,fixed zerofill, set thousands separator={},precision=0]{5.8092e1})&\pgfmathprintnumber [fixed,fixed zerofill,precision=4]{1.343e0}(\pgfmathprintnumber [fixed,fixed zerofill, set thousands separator={},precision=0]{5.2713e1})\\%
&\pgfutilensuremath {5}&\pgfutilensuremath {0.4657}&\pgfmathprintnumber [fixed,fixed zerofill,precision=4]{3.7271e-1}(\pgfmathprintnumber [fixed,fixed zerofill, set thousands separator={},precision=0]{2.4234e1})&\pgfmathprintnumber [fixed,fixed zerofill,precision=4]{1.2364e0}(\pgfmathprintnumber [fixed,fixed zerofill, set thousands separator={},precision=0]{6.1102e1})&\pgfmathprintnumber [fixed,fixed zerofill,precision=4]{1.4102e0}(\pgfmathprintnumber [fixed,fixed zerofill, set thousands separator={},precision=0]{5.399e1})\\%
&\pgfutilensuremath {6}&\pgfutilensuremath {0.5589}&\pgfmathprintnumber [fixed,fixed zerofill,precision=4]{4.9228e-1}(\pgfmathprintnumber [fixed,fixed zerofill, set thousands separator={},precision=0]{2.5554e1})&\pgfmathprintnumber [fixed,fixed zerofill,precision=4]{1.2654e0}(\pgfmathprintnumber [fixed,fixed zerofill, set thousands separator={},precision=0]{6.6136e1})&\pgfmathprintnumber [fixed,fixed zerofill,precision=4]{1.4822e0}(\pgfmathprintnumber [fixed,fixed zerofill, set thousands separator={},precision=0]{5.1996e1})\\%
&\pgfutilensuremath {7}&\pgfutilensuremath {0.6520}&\pgfmathprintnumber [fixed,fixed zerofill,precision=4]{6.0585e-1}(\pgfmathprintnumber [fixed,fixed zerofill, set thousands separator={},precision=0]{2.802e1})&\pgfmathprintnumber [fixed,fixed zerofill,precision=4]{1.3045e0}(\pgfmathprintnumber [fixed,fixed zerofill, set thousands separator={},precision=0]{7.1785e1})&\pgfmathprintnumber [fixed,fixed zerofill,precision=4]{1.5588e0}(\pgfmathprintnumber [fixed,fixed zerofill, set thousands separator={},precision=0]{5.2405e1})\\%
&\pgfutilensuremath {8}&\pgfutilensuremath {0.7452}&\pgfmathprintnumber [fixed,fixed zerofill,precision=4]{7.1588e-1}(\pgfmathprintnumber [fixed,fixed zerofill, set thousands separator={},precision=0]{3.1839e1})&\pgfmathprintnumber [fixed,fixed zerofill,precision=4]{1.3538e0}(\pgfmathprintnumber [fixed,fixed zerofill, set thousands separator={},precision=0]{7.8857e1})&\pgfmathprintnumber [fixed,fixed zerofill,precision=4]{1.6467e0}(\pgfmathprintnumber [fixed,fixed zerofill, set thousands separator={},precision=0]{5.293e1})\\%
&\pgfutilensuremath {9}&\pgfutilensuremath {0.8383}&\pgfmathprintnumber [fixed,fixed zerofill,precision=4]{8.2396e-1}(\pgfmathprintnumber [fixed,fixed zerofill, set thousands separator={},precision=0]{3.7134e1})&\pgfmathprintnumber [fixed,fixed zerofill,precision=4]{1.4148e0}(\pgfmathprintnumber [fixed,fixed zerofill, set thousands separator={},precision=0]{7.4219e1})&\pgfmathprintnumber [fixed,fixed zerofill,precision=4]{1.7273e0}(\pgfmathprintnumber [fixed,fixed zerofill, set thousands separator={},precision=0]{5.3876e1})\\%
&\pgfutilensuremath {10}&\pgfutilensuremath {0.9315}&\pgfmathprintnumber [fixed,fixed zerofill,precision=4]{9.3082e-1}(\pgfmathprintnumber [fixed,fixed zerofill, set thousands separator={},precision=0]{4.4415e1})&\pgfmathprintnumber [fixed,fixed zerofill,precision=4]{1.4757e0}(\pgfmathprintnumber [fixed,fixed zerofill, set thousands separator={},precision=0]{7.8899e1})&\pgfmathprintnumber [fixed,fixed zerofill,precision=4]{1.8153e0}(\pgfmathprintnumber [fixed,fixed zerofill, set thousands separator={},precision=0]{5.5386e1})\\%
&\pgfutilensuremath {11}&\pgfutilensuremath {1.0246}&\pgfmathprintnumber [fixed,fixed zerofill,precision=4]{1.03505e0}(\pgfmathprintnumber [fixed,fixed zerofill, set thousands separator={},precision=0]{5.4332e1})&\pgfmathprintnumber [fixed,fixed zerofill,precision=4]{1.5407e0}(\pgfmathprintnumber [fixed,fixed zerofill, set thousands separator={},precision=0]{8.454e1})&\pgfmathprintnumber [fixed,fixed zerofill,precision=4]{1.9104e0}(\pgfmathprintnumber [fixed,fixed zerofill, set thousands separator={},precision=0]{5.7515e1})\\%
&\pgfutilensuremath {12}&\pgfutilensuremath {1.1178}&\pgfmathprintnumber [fixed,fixed zerofill,precision=4]{1.1345e0}(\pgfmathprintnumber [fixed,fixed zerofill, set thousands separator={},precision=0]{6.8694e1})&\pgfmathprintnumber [fixed,fixed zerofill,precision=4]{1.609e0}(\pgfmathprintnumber [fixed,fixed zerofill, set thousands separator={},precision=0]{9.1087e1})&\pgfmathprintnumber [fixed,fixed zerofill,precision=4]{2.0026e0}(\pgfmathprintnumber [fixed,fixed zerofill, set thousands separator={},precision=0]{6.0503e1})\\\bottomrule %
\end {tabular}%

			\caption{Lattice data points defined in Eqs.\ (\ref{eq:def_V_Sigmagplus_tilde}) and (\ref{eq:def_V_hybrid_tilde}), where the self energy as well as lattice discretization errors at tree-level and proportional to $a^2$ are removed (using Fit~1), for the $\Sigma_g^+$, $\Pi_u$ and $\Sigma_u^-$ static potentials in units of $\text{GeV}$ (physical units are introduced by setting $r_0 = 0.5 \, \text{fm}$).}
			\label{tab:latticedata_ABCD_tilde_all}
		\end{table}
		\renewcommand{\arraystretch}{1.0}

	\section{$\text{SU(2)}$ lattice field theory results for the $\Sigma_g^+$, $\Pi_u$ and $\Sigma_u^-$ static potentials} \label{Appendix:SU2results}

		We carried out computations for gauge group $\text{SU(2)}$ analogous to those for gauge group $\text{SU(3)}$ discussed and presented in the main sections of this work.
		We generated three ensembles of gauge link configurations with gauge couplings $\beta = 2.85, \, 2.70, \, 2.50$. 
		We relate the lattice spacing $a$ to the scale $t_0$ using a parametrization of $\ln(t_0/a^2)$ determined in Ref.\ \cite{Hirakida:2018uoy} via the gradient flow. 
		Physical units are then introduced by setting $\sqrt{8 t_0} = 0.3010 \, \text{fm}$, which corresponds to $r_0 = 0.5 \, \text{fm}$.
		The details of the gauge link ensembles are summarized in Table~\ref{tab:latticesetupsSU2}.
		The three lattice volumes are quite similar, $L^3 \times T \approx (1.3 \, \text{fm})^3 \times (1.3 \, \text{fm})$.
		For the investigation of finite volume effects in Section~\ref{sec:finitevolume}, additional ensembles with both smaller and larger lattice volumes at gauge couplings $\beta= 3.00 , \, 2.85 , \, 2.70 , \, 2.50$ were generated. 
		
		\begin{table}[htb]
			\begin{center}
				\def\arraystretch{1.2}
				\begin{tabular}{cccccccccc}
					\hline
					ensemble & $\beta$ & $a$ in $\text{fm}$ \cite{Hirakida:2018uoy} & $(L/a)^3 \times T/a$ & $N_{\text{sim}}$ & $N_{\text{total}}$ & $N_{\text{or}}$  & $N_{\text{therm}}$ & $N_{\text{sep}}$ & $N_{\text{meas}}$ \\
					\hline
					$a$ & $2.50$ & $0.078$ & $16^3\times 16$ & $20$ & 
					$40 000$ & $0$ & $10 000$ & $100$ & $6000$ \\
					$b$ & $2.70$ & $0.041$ & $32^3\times 32$ & $20$ & 
					$25 000$ & $0$ & $10 000$ & $100$ & $3000$ \\
					$c$ & $2.85$ & $0.026$ & $48^3\times 48$ & $20$ & 
					$25 000$ & $0$ & $10 000$ & $200$ & $1500$ \\ 
					\hline
				\end{tabular}
			\end{center}
			\caption{Gauge link ensembles for gauge group $\text{SU(2)}$.}
			\label{tab:latticesetupsSU2}
		\end{table}
		
		\begin{table}[h]\centering
			\begin {tabular}{ccccc}%
\toprule ensemble&$r/a$&$V^e_{\Sigma _g^+ }\, a$&$V^e_{\Pi _u }\, a$&$V^e_{\Sigma _u^- }\, a$\\\midrule %
\multirow {7}{*}{$a$}&\pgfutilensuremath {2}&\pgfmathprintnumber [fixed,fixed zerofill,precision=6]{4.84307714000e-01}(\pgfmathprintnumber [fixed,fixed zerofill, set thousands separator={},precision=0]{6.8488678e1})&\pgfmathprintnumber [fixed,fixed zerofill,precision=4]{1.11504881239e+00}(\pgfmathprintnumber [fixed,fixed zerofill,precision=0]{4.6961166e1})&\pgfmathprintnumber [fixed,fixed zerofill,precision=4]{1.14482158600e+00}(\pgfmathprintnumber [fixed,fixed zerofill,precision=0]{5.7527863e1})\\%
&\pgfutilensuremath {3}&\pgfmathprintnumber [fixed,fixed zerofill,precision=6]{5.65441927000e-01}(\pgfmathprintnumber [fixed,fixed zerofill, set thousands separator={},precision=0]{1.511229e2})&\pgfmathprintnumber [fixed,fixed zerofill,precision=4]{1.08992692175e+00}(\pgfmathprintnumber [fixed,fixed zerofill,precision=0]{4.2160217e1})&\pgfmathprintnumber [fixed,fixed zerofill,precision=4]{1.14280357582e+00}(\pgfmathprintnumber [fixed,fixed zerofill,precision=0]{6.0667221e1})\\%
&\pgfutilensuremath {4}&\pgfmathprintnumber [fixed,fixed zerofill,precision=6]{6.23418081000e-01}(\pgfmathprintnumber [fixed,fixed zerofill, set thousands separator={},precision=0]{2.7072632e2})&\pgfmathprintnumber [fixed,fixed zerofill,precision=4]{1.08030883741e+00}(\pgfmathprintnumber [fixed,fixed zerofill,precision=0]{3.800882e1})&\pgfmathprintnumber [fixed,fixed zerofill,precision=4]{1.14859044592e+00}(\pgfmathprintnumber [fixed,fixed zerofill,precision=0]{6.6287735e1})\\%
&\pgfutilensuremath {5}&\pgfmathprintnumber [fixed,fixed zerofill,precision=6]{6.71505549000e-01}(\pgfmathprintnumber [fixed,fixed zerofill, set thousands separator={},precision=0]{4.2249039e2})&\pgfmathprintnumber [fixed,fixed zerofill,precision=4]{1.07921282212e+00}(\pgfmathprintnumber [fixed,fixed zerofill,precision=0]{3.9233551e1})&\pgfmathprintnumber [fixed,fixed zerofill,precision=4]{1.16769085251e+00}(\pgfmathprintnumber [fixed,fixed zerofill,precision=0]{7.3904831e1})\\%
&\pgfutilensuremath {6}&\pgfmathprintnumber [fixed,fixed zerofill,precision=6]{7.14737013000e-01}(\pgfmathprintnumber [fixed,fixed zerofill, set thousands separator={},precision=0]{6.4498474e2})&\pgfmathprintnumber [fixed,fixed zerofill,precision=4]{1.08189170777e+00}(\pgfmathprintnumber [fixed,fixed zerofill,precision=0]{3.9283173e1})&\pgfmathprintnumber [fixed,fixed zerofill,precision=4]{1.18643834529e+00}(\pgfmathprintnumber [fixed,fixed zerofill,precision=0]{8.6147446e1})\\%
&\pgfutilensuremath {7}&\pgfmathprintnumber [fixed,fixed zerofill,precision=6]{7.55318480000e-01}(\pgfmathprintnumber [fixed,fixed zerofill, set thousands separator={},precision=0]{9.1388275e2})&\pgfmathprintnumber [fixed,fixed zerofill,precision=4]{1.09010012426e+00}(\pgfmathprintnumber [fixed,fixed zerofill,precision=0]{4.261e1})&\pgfmathprintnumber [fixed,fixed zerofill,precision=4]{1.21491265129e+00}(\pgfmathprintnumber [fixed,fixed zerofill,precision=0]{9.7942459e1})\\%
&\pgfutilensuremath {8}&\pgfmathprintnumber [fixed,fixed zerofill,precision=6]{7.94584025000e-01}(\pgfmathprintnumber [fixed,fixed zerofill, set thousands separator={},precision=0]{1.2923828e3})&\pgfmathprintnumber [fixed,fixed zerofill,precision=4]{1.10117155216e+00}(\pgfmathprintnumber [fixed,fixed zerofill,precision=0]{4.587851e1})&\pgfmathprintnumber [fixed,fixed zerofill,precision=4]{1.23072483715e+00}(\pgfmathprintnumber [fixed,fixed zerofill,precision=0]{1.1498062e2})\\%
\midrule \multirow {9}{*}{$b$}&\pgfutilensuremath {2}&\pgfmathprintnumber [fixed,fixed zerofill,precision=6]{3.95714346957e-01}(\pgfmathprintnumber [fixed,fixed zerofill, set thousands separator={},precision=0]{7.1189377e0})&\pgfmathprintnumber [fixed,fixed zerofill,precision=4]{8.21330962490e-01}(\pgfmathprintnumber [fixed,fixed zerofill,precision=0]{1.7556015e1})&\pgfmathprintnumber [fixed,fixed zerofill,precision=4]{8.25891510388e-01}(\pgfmathprintnumber [fixed,fixed zerofill,precision=0]{1.9443237e1})\\%
&\pgfutilensuremath {3}&\pgfmathprintnumber [fixed,fixed zerofill,precision=6]{4.46300180526e-01}(\pgfmathprintnumber [fixed,fixed zerofill, set thousands separator={},precision=0]{1.6697876e1})&\pgfmathprintnumber [fixed,fixed zerofill,precision=4]{8.03958845654e-01}(\pgfmathprintnumber [fixed,fixed zerofill,precision=0]{2.9284882e1})&\pgfmathprintnumber [fixed,fixed zerofill,precision=4]{8.14371669767e-01}(\pgfmathprintnumber [fixed,fixed zerofill,precision=0]{1.985762e1})\\%
&\pgfutilensuremath {4}&\pgfmathprintnumber [fixed,fixed zerofill,precision=6]{4.76883019032e-01}(\pgfmathprintnumber [fixed,fixed zerofill, set thousands separator={},precision=0]{3.0979446e1})&\pgfmathprintnumber [fixed,fixed zerofill,precision=4]{7.96384344191e-01}(\pgfmathprintnumber [fixed,fixed zerofill,precision=0]{1.6477234e1})&\pgfmathprintnumber [fixed,fixed zerofill,precision=4]{8.10402742562e-01}(\pgfmathprintnumber [fixed,fixed zerofill,precision=0]{2.0447845e1})\\%
&\pgfutilensuremath {5}&\pgfmathprintnumber [fixed,fixed zerofill,precision=6]{4.99215850500e-01}(\pgfmathprintnumber [fixed,fixed zerofill, set thousands separator={},precision=0]{5.0273834e1})&\pgfmathprintnumber [fixed,fixed zerofill,precision=4]{7.90250419645e-01}(\pgfmathprintnumber [fixed,fixed zerofill,precision=0]{1.602031e1})&\pgfmathprintnumber [fixed,fixed zerofill,precision=4]{8.10135003625e-01}(\pgfmathprintnumber [fixed,fixed zerofill,precision=0]{2.1098312e1})\\%
&\pgfutilensuremath {6}&\pgfmathprintnumber [fixed,fixed zerofill,precision=6]{5.17411408871e-01}(\pgfmathprintnumber [fixed,fixed zerofill, set thousands separator={},precision=0]{7.4140594e1})&\pgfmathprintnumber [fixed,fixed zerofill,precision=4]{7.86000207800e-01}(\pgfmathprintnumber [fixed,fixed zerofill,precision=0]{1.5576889e1})&\pgfmathprintnumber [fixed,fixed zerofill,precision=4]{8.11805946011e-01}(\pgfmathprintnumber [fixed,fixed zerofill,precision=0]{2.1996277e1})\\%
&\pgfutilensuremath {7}&\pgfmathprintnumber [fixed,fixed zerofill,precision=6]{5.33190752517e-01}(\pgfmathprintnumber [fixed,fixed zerofill, set thousands separator={},precision=0]{1.0350601e2})&\pgfmathprintnumber [fixed,fixed zerofill,precision=4]{7.83159716981e-01}(\pgfmathprintnumber [fixed,fixed zerofill,precision=0]{1.5342987e1})&\pgfmathprintnumber [fixed,fixed zerofill,precision=4]{8.15316891693e-01}(\pgfmathprintnumber [fixed,fixed zerofill,precision=0]{2.2953842e1})\\%
&\pgfutilensuremath {8}&\pgfmathprintnumber [fixed,fixed zerofill,precision=6]{5.47434884464e-01}(\pgfmathprintnumber [fixed,fixed zerofill, set thousands separator={},precision=0]{1.3768036e2})&\pgfmathprintnumber [fixed,fixed zerofill,precision=4]{7.81697907126e-01}(\pgfmathprintnumber [fixed,fixed zerofill,precision=0]{1.5240189e1})&\pgfmathprintnumber [fixed,fixed zerofill,precision=4]{8.19766927984e-01}(\pgfmathprintnumber [fixed,fixed zerofill,precision=0]{2.4193039e1})\\%
&\pgfutilensuremath {9}&\pgfmathprintnumber [fixed,fixed zerofill,precision=6]{5.60629647338e-01}(\pgfmathprintnumber [fixed,fixed zerofill, set thousands separator={},precision=0]{1.7742737e2})&\pgfmathprintnumber [fixed,fixed zerofill,precision=4]{7.80747790095e-01}(\pgfmathprintnumber [fixed,fixed zerofill,precision=0]{1.5278946e1})&\pgfmathprintnumber [fixed,fixed zerofill,precision=4]{8.25413617230e-01}(\pgfmathprintnumber [fixed,fixed zerofill,precision=0]{2.5451797e1})\\%
&\pgfutilensuremath {10}&\pgfmathprintnumber [fixed,fixed zerofill,precision=6]{5.73100949545e-01}(\pgfmathprintnumber [fixed,fixed zerofill, set thousands separator={},precision=0]{2.2331711e2})&\pgfmathprintnumber [fixed,fixed zerofill,precision=4]{7.81176221245e-01}(\pgfmathprintnumber [fixed,fixed zerofill,precision=0]{1.5484131e1})&\pgfmathprintnumber [fixed,fixed zerofill,precision=4]{8.31391917397e-01}(\pgfmathprintnumber [fixed,fixed zerofill,precision=0]{2.6962341e1})\\%
\midrule \multirow {10}{*}{$c$}&\pgfutilensuremath {2}&\pgfmathprintnumber [fixed,fixed zerofill,precision=6]{3.53847768741e-01}(\pgfmathprintnumber [fixed,fixed zerofill, set thousands separator={},precision=0]{1.6395813e1})&\pgfmathprintnumber [fixed,fixed zerofill,precision=4]{6.85997017126e-01}(\pgfmathprintnumber [fixed,fixed zerofill,precision=0]{2.5153687e1})&\pgfmathprintnumber [fixed,fixed zerofill,precision=4]{6.85436874039e-01}(\pgfmathprintnumber [fixed,fixed zerofill,precision=0]{2.3993347e1})\\%
&\pgfutilensuremath {3}&\pgfmathprintnumber [fixed,fixed zerofill,precision=6]{3.94721769269e-01}(\pgfmathprintnumber [fixed,fixed zerofill, set thousands separator={},precision=0]{2.9733887e1})&\pgfmathprintnumber [fixed,fixed zerofill,precision=4]{6.73434170579e-01}(\pgfmathprintnumber [fixed,fixed zerofill,precision=0]{2.2697708e1})&\pgfmathprintnumber [fixed,fixed zerofill,precision=4]{6.73738252928e-01}(\pgfmathprintnumber [fixed,fixed zerofill,precision=0]{2.2668594e1})\\%
&\pgfutilensuremath {4}&\pgfmathprintnumber [fixed,fixed zerofill,precision=6]{4.17829155647e-01}(\pgfmathprintnumber [fixed,fixed zerofill, set thousands separator={},precision=0]{5.0349533e1})&\pgfmathprintnumber [fixed,fixed zerofill,precision=4]{6.64596220906e-01}(\pgfmathprintnumber [fixed,fixed zerofill,precision=0]{2.1909775e1})&\pgfmathprintnumber [fixed,fixed zerofill,precision=4]{6.67953596422e-01}(\pgfmathprintnumber [fixed,fixed zerofill,precision=0]{2.139856e1})\\%
&\pgfutilensuremath {5}&\pgfmathprintnumber [fixed,fixed zerofill,precision=6]{4.33610824865e-01}(\pgfmathprintnumber [fixed,fixed zerofill, set thousands separator={},precision=0]{8.7234344e1})&\pgfmathprintnumber [fixed,fixed zerofill,precision=4]{6.60921461300e-01}(\pgfmathprintnumber [fixed,fixed zerofill,precision=0]{2.1986267e1})&\pgfmathprintnumber [fixed,fixed zerofill,precision=4]{6.65449097224e-01}(\pgfmathprintnumber [fixed,fixed zerofill,precision=0]{2.2413055e1})\\%
&\pgfutilensuremath {6}&\pgfmathprintnumber [fixed,fixed zerofill,precision=6]{4.45684613536e-01}(\pgfmathprintnumber [fixed,fixed zerofill, set thousands separator={},precision=0]{8.9201523e1})&\pgfmathprintnumber [fixed,fixed zerofill,precision=4]{6.56744804230e-01}(\pgfmathprintnumber [fixed,fixed zerofill,precision=0]{1.9339264e1})&\pgfmathprintnumber [fixed,fixed zerofill,precision=4]{6.64286914251e-01}(\pgfmathprintnumber [fixed,fixed zerofill,precision=0]{2.195021e1})\\%
&\pgfutilensuremath {7}&\pgfmathprintnumber [fixed,fixed zerofill,precision=6]{4.55556957370e-01}(\pgfmathprintnumber [fixed,fixed zerofill, set thousands separator={},precision=0]{1.1032623e2})&\pgfmathprintnumber [fixed,fixed zerofill,precision=4]{6.52416807334e-01}(\pgfmathprintnumber [fixed,fixed zerofill,precision=0]{1.8561539e1})&\pgfmathprintnumber [fixed,fixed zerofill,precision=4]{6.64408281648e-01}(\pgfmathprintnumber [fixed,fixed zerofill,precision=0]{2.2234283e1})\\%
&\pgfutilensuremath {8}&\pgfmathprintnumber [fixed,fixed zerofill,precision=6]{4.63982091713e-01}(\pgfmathprintnumber [fixed,fixed zerofill, set thousands separator={},precision=0]{1.4054474e2})&\pgfmathprintnumber [fixed,fixed zerofill,precision=4]{6.50327573393e-01}(\pgfmathprintnumber [fixed,fixed zerofill,precision=0]{1.7853516e1})&\pgfmathprintnumber [fixed,fixed zerofill,precision=4]{6.64348835448e-01}(\pgfmathprintnumber [fixed,fixed zerofill,precision=0]{2.2994705e1})\\%
&\pgfutilensuremath {9}&\pgfmathprintnumber [fixed,fixed zerofill,precision=6]{4.71474297201e-01}(\pgfmathprintnumber [fixed,fixed zerofill, set thousands separator={},precision=0]{1.644365e2})&\pgfmathprintnumber [fixed,fixed zerofill,precision=4]{6.49303987171e-01}(\pgfmathprintnumber [fixed,fixed zerofill,precision=0]{1.926886e1})&\pgfmathprintnumber [fixed,fixed zerofill,precision=4]{6.63473776984e-01}(\pgfmathprintnumber [fixed,fixed zerofill,precision=0]{2.48284e1})\\%
&\pgfutilensuremath {10}&\pgfmathprintnumber [fixed,fixed zerofill,precision=6]{4.78363928411e-01}(\pgfmathprintnumber [fixed,fixed zerofill, set thousands separator={},precision=0]{2.1418289e2})&\pgfmathprintnumber [fixed,fixed zerofill,precision=4]{6.45802172271e-01}(\pgfmathprintnumber [fixed,fixed zerofill,precision=0]{1.8376572e1})&\pgfmathprintnumber [fixed,fixed zerofill,precision=4]{6.64825604539e-01}(\pgfmathprintnumber [fixed,fixed zerofill,precision=0]{2.5949768e1})\\%
&\pgfutilensuremath {11}&\pgfmathprintnumber [fixed,fixed zerofill,precision=6]{4.84683678141e-01}(\pgfmathprintnumber [fixed,fixed zerofill, set thousands separator={},precision=0]{2.2834702e2})&\pgfmathprintnumber [fixed,fixed zerofill,precision=4]{6.43794599820e-01}(\pgfmathprintnumber [fixed,fixed zerofill,precision=0]{1.928566e1})&\pgfmathprintnumber [fixed,fixed zerofill,precision=4]{6.66207178730e-01}(\pgfmathprintnumber [fixed,fixed zerofill,precision=0]{2.6899643e1})\\\bottomrule %
\end {tabular}%

			\caption{Bare lattice data points for the $\Sigma_g^+$, $\Pi_u$ and $\Sigma_u^-$ static potentials in units of the lattice spacing (see Table~\ref{tab:latticesetupsSU2}) for gauge group $\text{SU(2)}$.}
			\label{tab:latticedata_SU2_all}
		\end{table}
		
		In Table~\ref{tab:latticedata_SU2_all} we list $V^e_{\Lambda_{\eta}^{\epsilon}}(r)a$, the bare lattice data points in units of the lattice spacing. 
		These can be used to generate parametrizations, using methods as e.g.\ discussed in Section~\ref{sec:parametrization}.
	\clearpage


\bibliographystyle{utphys.bst}
\bibliography{hybrid_potentials}

\end{document}